\documentclass[apj]{emulateapj}

\usepackage{graphicx}
\usepackage{txfonts}
\usepackage{array}
\usepackage{amssymb}
\usepackage{natbib}
\usepackage{lscape}
\usepackage{rotating}
\usepackage{longtable}

\newcommand{\hd}{HD\,5980}
\newcommand{\xmm}{{\sc{XMM}}\emph{-Newton}}
\newcommand{\ch}{\emph{Chandra}}
\shorttitle{A changing wind collision}
\shortauthors{Naz\'e et al.}

\begin{document}

\title{A changing wind collision\footnote{Based on \xmm\ and \ch\ data.}}

\author{Ya\"el Naz\'e\altaffilmark{1,2}}
\affil{Groupe d'Astrophysique des Hautes Energies, STAR, Universit\'e de Li\`ege, Quartier Agora (B5c, Institut d'Astrophysique et de G\'eophysique), All\'ee du 6 Ao\^ut 19c, B-4000 Sart Tilman, Li\`ege, Belgium}
\and
\author{Gloria Koenigsberger}
\affil{Instituto de Ciencias F\'{\i}sicas, Universidad Nacional Aut\'onoma de M\'exico, Ave. Universidad S/N, Cuernavaca, 62210, Morelos, M\'exico}
\and
\author{Julian M. Pittard}
\affil{School of Physics and Astronomy, University of Leeds, Woodhouse Lane, Leeds LS2 9JT, UK}
\and
\author{Elliot Ross Parkin}
\affil{Groupe d'Astrophysique des Hautes Energies, STAR, Universit\'e de Li\`ege, Quartier Agora (B5c, Institut d'Astrophysique et de G\'eophysique), All\'ee du 6 Ao\^ut 19c, B-4000 Sart Tilman, Li\`ege, Belgium}
\and
\author{Gregor Rauw}
\affil{Groupe d'Astrophysique des Hautes Energies, STAR, Universit\'e de Li\`ege, Quartier Agora (B5c, Institut d'Astrophysique et de G\'eophysique), All\'ee du 6 Ao\^ut 19c, B-4000 Sart Tilman, Li\`ege, Belgium}
\and
\author{Michael F. Corcoran}
\affil{CRESST and X-ray Astrophysics Laboratory, NASA/Goddard Space Flight Center, Greenbelt, MD 20771, USA ; The Catholic University of America, 620 Michigan Ave., N.E. Washington, DC 20064, USA}
\and
\author{D. John Hillier}
\affil{Department of Physics and Astronomy \& Pittsburgh Particle Physics, Astrophysics and Cosmology Center (PITT PACC), University of Pittsburgh, Pittsburgh, PA 15260, USA}

\altaffiltext{1}{F.R.S.-FNRS Research Associate.}
\altaffiltext{2}{naze@astro.ulg.ac.be}

\begin{abstract}
We report on the first detection of a global change in the X-ray emitting properties of a wind-wind collision, thanks to \xmm\ observations of the massive SMC system \hd.  While its lightcurve had remained unchanged between 2000 and 2005, the X-ray flux has now increased by a factor of $\sim$2.5, and slightly hardened. The new observations also extend the observational coverage over the entire orbit, pinpointing the lightcurve shape. It has not varied much despite the large overall brightening, and a tight correlation of fluxes with orbital separation is found, without any hysteresis effect. Moreover, the absence of eclipses and of absorption effects related to orientation suggests a large size for the X-ray emitting region. Simple analytical models of the wind-wind collision, considering the varying wind properties of the eruptive component in \hd, are able to reproduce the recent hardening and the flux-separation relationship, at least qualitatively, but they predict a hardening at apastron and little change in mean flux, contrary to observations. The brightness change could then possibly be related to a recently theorized phenomenon linked to the varying strength of thin-shell instabilities in shocked wind regions.
\end{abstract}


\section{Introduction} 

Luminous Blue Variables (LBVs) represent a short-lived but important stage in the life of the most massive stars. The massive ejection of material at this stage has a great impact both on the star's fate, as the stellar mass drastically changes, and on the local interstellar environment, as large amounts of kinetic energy as well as chemical elements are injected in the surroundings. However, our knowledge of LBVs remains incomplete. The cause of these eruptions, their recurrence timescales, or their effect on the subsequent evolution of the star are still not fully understood. The most recent such eruption in our Galaxy was the famous ``Great Eruption'' in the mid-19th century of $\eta$\,Carinae, in which the kinetic energy of the ejected material rivaled the energy output of a weak supernova. $\eta$\,Carinae has a companion whose strong stellar wind collides with that of the eruptor, and such collisions provide interesting constraints on wind properties. However, for $\eta$\,Carinae, it has been challenging to deduce from historical records how the interaction has evolved over time. A unique opportunity to overcome this challenge occurred in 1994 with the eruption of the primary star in \hd, the most luminous multiple star system in the Small Magellanic Cloud (SMC). 

The \hd\ system lies on the outskirts of NGC\,346. It consists of several stars. Star A, initially thought to have been an O-type supergiant \citep{bre82}, is the component that underwent a major LBV eruption in 1994. Its optical brightness has been gradually declining since then \citep{koe10}. The post-eruption spectrum of Star A transitioned from WN11 to WN6, which brings into question the original classification of this star as an O-type supergiant, rather favoring a WR classification \citep{nie88,koe14}. Star A possesses a WR companion called Star B, with which it forms an eclipsing pair \citep[$P_{A+B}$=19.3\,d, $i$=86$^{\circ}$, ][]{ste97,per09}. The presence of a third star, adequately named Star C, was detected through its contribution (about 30\%) to the optical lightcurve \citep{ste97}. Its presence was confirmed in optical/UV spectroscopy as its photospheric absorption lines appear superposed on the WR spectra (which may have confused the original classification of Star A). These lines display velocity variations with $P_C=96.5d$ \citep[and references therein]{sch00,foe08,koe14}. This period is too short for Star C to be in a stable orbit around the A+B pair \citep[e.g.][]{tok04}, hence Star C is itself part of a binary, which either serendipitously lies along the line-of-sight of the A+B pair or orbits around it with a very long period.  

The emission lines in \hd\ undergo line profile variations with $P_{A+B}$, which were attributed to colliding winds by \citet{mof98}. It has however been argued that eclipse effects and an asymmetric wind of Star A provide alternative explanations for this variability \citep{koe06}. Nevertheless, the presence of two powerful winds in such a massive binary system should lead to their interaction. \citet{foe08} speculated on the geometry of the interacting region, based on the variability that was observed in the blue-shifted absorptions in the optical He\,{\sc i} lines. They suggested that, at distances from the A+B binary larger than the orbital separation, the interaction could be described as two spiral-like density enhancements which, when projected against the stellar continua, produced the blue-shifted absorptions. Direct evidence for colliding winds, closer to the stars, was obtained in the X-ray range: \hd\ appears as a very bright and hard X-ray source \citep{naz02}, two features typical of colliding winds binaries, and phase-locked variations were also detected thanks to a dedicated monitoring \citep{naz07}.

Since those observations, the wind velocity of Star A has further increased, accompanied by a further decrease in its mass-loss rate. Because the X-ray emission is sensitive to mass loss, \hd\ provides the opportunity to determine how changes in the mass loss from Star A affect the shocked plasma in a system with a known orbital configuration, a unique astrophysical experiment. We have therefore undertaken a new monitoring in 2016 with the \xmm\ observatory, to document the variations in the hot shocked gas with the changing strength of the wind from Star A. Section 2 presents the data used in this paper, Section 3 presents and discusses the high-energy results, and Section 4 concludes by summarizing our results.

\section{Data} 

\subsection{\ch}
\hd\ was observed twice with \ch\ (Table~\ref{journal}). The first dataset (ObsID=1881, PI Corcoran) was obtained in May 2001 with ACIS-I and we have analyzed it in detail in \citet{naz02,naz03}. We reprocessed these data using CIAO 4.8 (CALDB 4.7.0). The spectrum of \hd\ was then extracted (using {\it specextract}) in a circle of radius 5px (2.5'') centered on the Simbad coordinates of the system, using the surrounding annulus (with radii of 5 and 15px) as background region. The second exposure (ObsID=13773, PI Ballet), taken in February 2013, was aimed at observing the nearby supernova remnant (SNR) IKT\,16, but \hd\ serendipitously lies within the ACIS-S field-of-view. Because of the off-axis angle, however, its PSF is wider and somewhat distorted. After reprocessing, the spectrum of \hd\ was thus extracted in a circular region of larger radius (15px), using a larger background region (annulus with radii of 15 and 25px). We calculated specific detector response matrices (RMF and ARF) to obtain energy- and flux-calibrated spectra. Finally, these spectra were grouped to a minimum of 15 cts per bin prior to analysis.

\begin{table*}
\centering
\footnotesize
\caption{Journal of the X-ray observations. HJD correspond to dates at mid-exposure, and the corresponding phases $\phi$ were calculated using the ephemeris of \citet[$T_0$=2\,443\,158.705, $P_{A+B}$=19.2654\,d]{ste97}. Phases $\phi_C$ correspond to Star C orbit \citep[$T_0$=2\,451\,183.4, $P_C$=96.56\,d, see Table 7 of][]{koe14}. Exposure times correspond to on-axis values (for MOS1 and excluding flares, except for Rev. 0970, if \xmm). The \xmm\ count rates in the 1.5--10.\,keV band are provided in the last two columns (first one for the sum of MOS1 and MOS2 count rates, second one for pn count rates - note that the pn camera was switched off during Rev. 0970).}
\label{journal}
\begin{tabular}{lccccccc}
\hline\hline
ID/Rev & Date & $\Delta(t)$ & HJD & $\phi$ & $\phi_C$ &MOS1+2 ct. rate & pn ct. rate\\
   &      & (ks)        &     &        &          &\multicolumn{2}{c}{($10^{-3}$\,cts\,s$^{-1}$)} \\
\hline
\multicolumn{6}{l}{\ch} \\
1881  & 2001-05-15 & 98.7 & 2452045.171 & 0.266 & 0.92 \\
13773 & 2013-02-09 & 38.6 & 2456332.837 & 0.823 & 0.33 \\
\multicolumn{6}{l}{\xmm} \\
0157 & 2000-10-17 & 17.4 & 2451835.246 & 0.369 & 0.75 & 13.0$\pm$1.1  & 14.2$\pm$1.5   \\
0357 & 2001-11-21 & 26.7 & 2452234.643 & 0.100 & 0.89 & 3.95$\pm$0.54 & 5.29$\pm$0.75  \\
0970 & 2005-03-27 & 23.5 & 2453457.416 & 0.570 & 0.55 & 19.7$\pm$1.6  &                \\
1093 & 2005-11-27 & 17.3 & 2453701.862 & 0.259 & 0.08 & 8.09$\pm$0.85 & 16.0$\pm$1.3   \\
1094 & 2005-11-29 & 16.4 & 2453703.806 & 0.360 & 0.10 & 13.4$\pm$1.1  & 15.8$\pm$1.4   \\
1100 & 2005-12-11 & 13.3 & 2453716.124 & 0.999 & 0.23 & 4.23$\pm$0.75 & 5.23$\pm$1.03  \\
3073 & 2016-09-19 & 21.4 & 2457651.187 & 0.254 & 0.98 & 24.5$\pm$1.4  & 32.3$\pm$1.7   \\
3074 & 2016-09-21 & 26.2 & 2457653.263 & 0.362 & 0.00 & 31.9$\pm$1.4  & 42.9$\pm$1.8   \\
3086 & 2016-10-15 & 26.8 & 2457677.143 & 0.602 & 0.25 & 36.3$\pm$1.4  & 54.5$\pm$1.9   \\
3088 & 2016-10-19 & 21.5 & 2457681.003 & 0.802 & 0.29 & 28.0$\pm$1.7  & 37.3$\pm$2.4   \\
3090 & 2016-10-23 & 22.7 & 2457684.806 & 0.999 & 0.33 & 13.1$\pm$1.0  & 14.3$\pm$1.1   \\
3091 & 2016-10-25 & 13.9 & 2457686.771 & 0.101 & 0.35 & 12.5$\pm$1.2  & 12.8$\pm$1.3   \\
3110 & 2016-12-02 & 30.9 & 2457725.322 & 0.102 & 0.75 & 13.3$\pm$0.9  & 18.5$\pm$1.1   \\
\hline
\end{tabular}
\end{table*}

\subsection{\xmm}
\hd\ was observed 13 times by \xmm\ in 2000, 2001, 2005, and 2016 (PIs Bleeker, Corcoran, Parmar, Naz\'e, respectively). The detailed journal of observations is provided in Table~\ref{journal}. The first analyses of the 2000--2005 observations were reported in \citet{naz04,naz07}. All datasets were reprocessed using SAS v16.0.0 using calibration files available in Winter 2016--2017 and following the recommendations of the \xmm\ team\footnote{see \\ http://xmm.esac.esa.int/sas/current/documentation/threads/ }. 

\begin{figure}
\includegraphics[width=8.5cm]{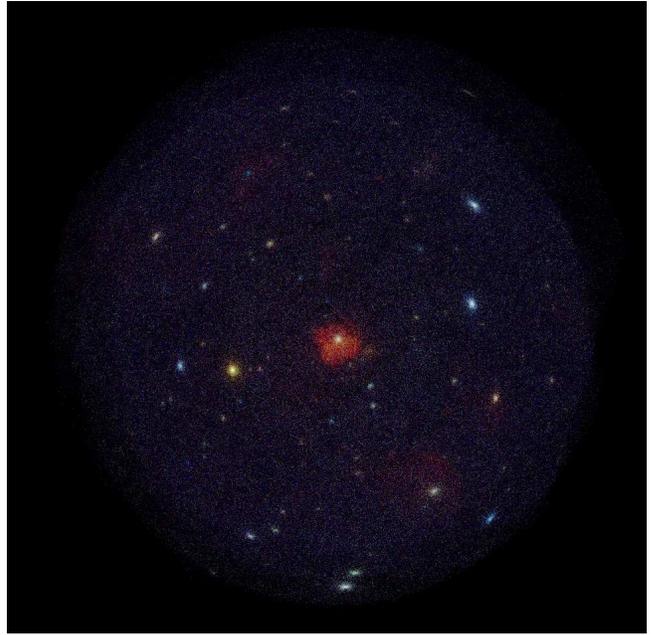}
\caption{EPIC image of \hd\ (the bright source at center) and its surroundings obtained by combining all \xmm\ observations. Red, green, and blue correspond to 0.5--1.5\,keV, 1.5--2.5\,keV and 2.5--10.\,keV energy bands, respectively. Note in particular the soft (red) emission surrounding \hd, which arises from the SNR IKT\,18. }
\label{color}
\end{figure}

\begin{figure}
\includegraphics[height=8.5cm, angle=270]{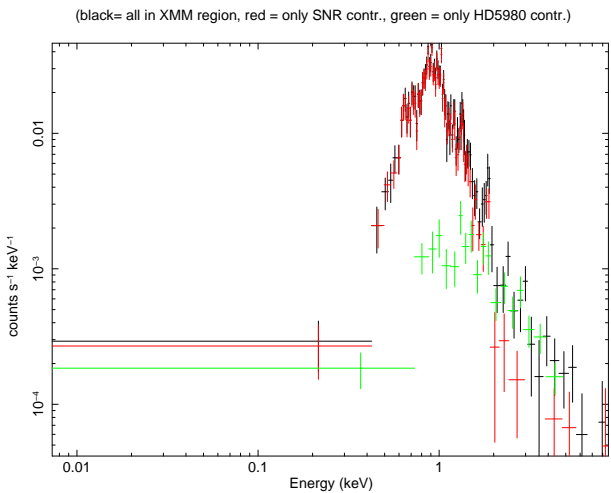}
\caption{The spectra recorded by \ch\ (ObsID 1881) for \hd\ only (in green), for \hd\ and part of the surrounding SNR (in black - 25'' circle centered on \hd, i.e., the region used for extracting \xmm\ spectra), and for the SNR contribution only (in red, annulus centered on \hd\ with radii 2.5'' and 25''). Note how the SNR contribution dominates over that of \hd\ in the \xmm\ region, especially below 2\,keV. }
\label{chspec}
\end{figure}

\begin{figure*}
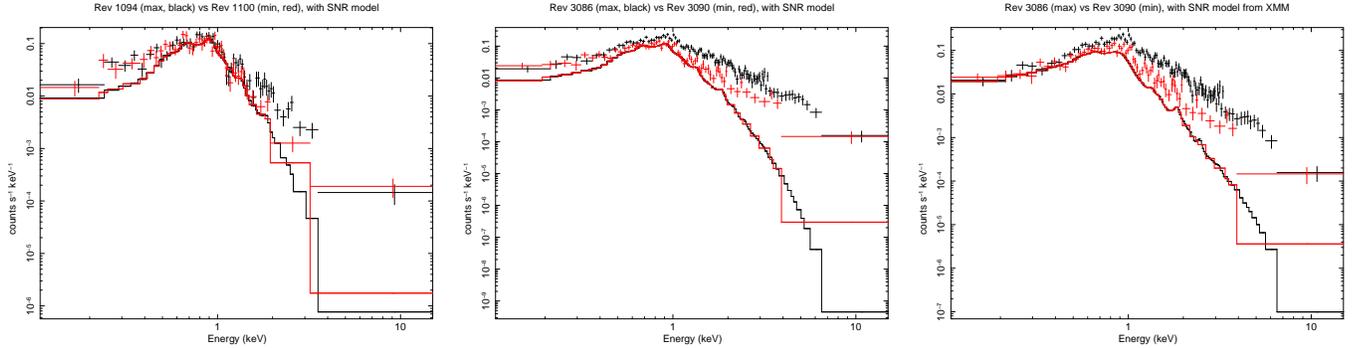

\includegraphics[height=6cm, angle=270]{newcompaxmm_minmax_2005.ps}
\includegraphics[height=6cm, angle=270]{newcompaxmm_minmax_2016.ps}
\includegraphics[height=6cm, angle=270]{newcompaxmm_minmax_2016b.ps}
\caption{\xmm\ spectra extracted over a circle of 25'' radius around \hd\ and considering an external background (i.e. the spectra combine a contribution from \hd\ and a contamination by the surrounding SNR emission). {\it Left and Middle:} EPIC-pn spectra corresponding to the minimum (red crosses) and maximum (black crosses) emissions of \hd\ in 2005 (left) and 2016 (middle), with the best-fit model (histogram-like solid lines) for the SNR contribution derived from \ch\ data in the same area. Note how the softest X-rays are not reproduced, despite being certainly due to the SNR. {\it Right:} EPIC-pn spectra corresponding to the minimum (red crosses) and maximum (black crosses) emissions in 2016, with the best-fit model (solid line) for the SNR contribution derived from \xmm\ data in a nearby area. The softest X-rays are now reproduced, showing that some cross-calibration problems remain between \ch-ACIS and \xmm-EPIC. }
\label{compaxmmchfit}
\end{figure*}

The EPIC observations, taken in full-frame mode and with the medium filter (to reject optical/UV light), were filtered to accept only the best-quality data ({\sc{pattern}} of 0--12 for MOS and 0--4 for pn). Background flares were detected for observations taken in Revs. 0157, 0970, 3074, 3086, 3088, 3090, 3091, and 3110. These data were thus filtered to eliminate times where the count rate above 10.\,keV was higher than 0.3\,cts\,s$^{-1}$ for MOS and 0.5\,cts\,s$^{-1}$ for pn. Unfortunately, the observation taken in Rev. 0970 was fully affected by flares: applying our flare filtering criteria results in no usable time. We have thus used these data without flare filtering but for source detection only. They were not included in the spectral analysis. Furthermore, the data taken in Rev. 1100 suffer from a slew failure at the very beginning of the observation, so we ignored the data during this interval. Finally, during Rev. 3091, there was a ground station outage during which no data could be recorded, but the data taken before and after this event were combined for further analyses.  

Since \hd\ is surrounded by the soft emission of the SNR IKT\,18 (Fig. \ref{color}), we performed the source detection, using the task {\it edetect\_chain} (for a log-likelihood of 10), only in the hard, 1.5--10\,keV, energy band. This was done in several steps, first using a sliding box and then point spread function (PSF) fitting: the final count rates correspond to equivalent on-axis, full PSF count rates. Table~\ref{journal} provides these EPIC count rates. 

We then extracted EPIC spectra of \hd\ using the task {\it{especget}}, with a binning ensuring a signal-to-noise ratio of at least 3 per bin; dedicated response matrices were calculated at the same time. The source region is a circle centered on the best-fit position determined by the source detection algorithm and with a radius of 25''. Since the PSF of \xmm\ is much broader than that of \ch, the emission of \hd\ here appears much more blended with the surrounding SNR (compare Fig. 1 of \citealt{naz04} with Fig. \ref{color} above). In addition, the X-ray emission in this region is generally dominated by the SNR, not \hd\ (Fig. \ref{chspec}). We have thus attempted several strategies to obtain a clean estimate of the spectral parameters of \hd. First, we have used as background the SNR itself, avoiding the pn gaps: the region used is a box with 92.8'' and 107.2'' dimensions whose center is offset from \hd\ position by 16.3'' east and 21.5'' south (i.e. -326.5\,px in X and -430\,px in Y), with the source region excised. This will correct for the SNR emission within the source region only if its emission is uniform, which is not exactly the case. Therefore, we also used a background region located outside the SNR, which is a circle of 25'' radius whose center is offset by 102'' east and 82'' south from \hd\ (i.e. -2040\,px in X and -1640\,px in Y, as in \citealt{naz07}). The derived background-corrected spectra were then fitted considering a (fixed) SNR contribution whose spectral parameters can be derived with \ch\ in the exact same region but after excising \hd. This procedure corrects for the SNR contribution in the exact surroundings of \hd, but it assumes a perfect cross-calibration between observatories, which is not exactly the case either (see Fig. \ref{compaxmmchfit}). Neither solution thus is perfect, but results from both methods are in fact similar, as further discussed in the next section.

\section{The wind-wind collision in \hd} 

\subsection{The X-ray lightcurve}

\begin{figure*}
\includegraphics[width=6cm]{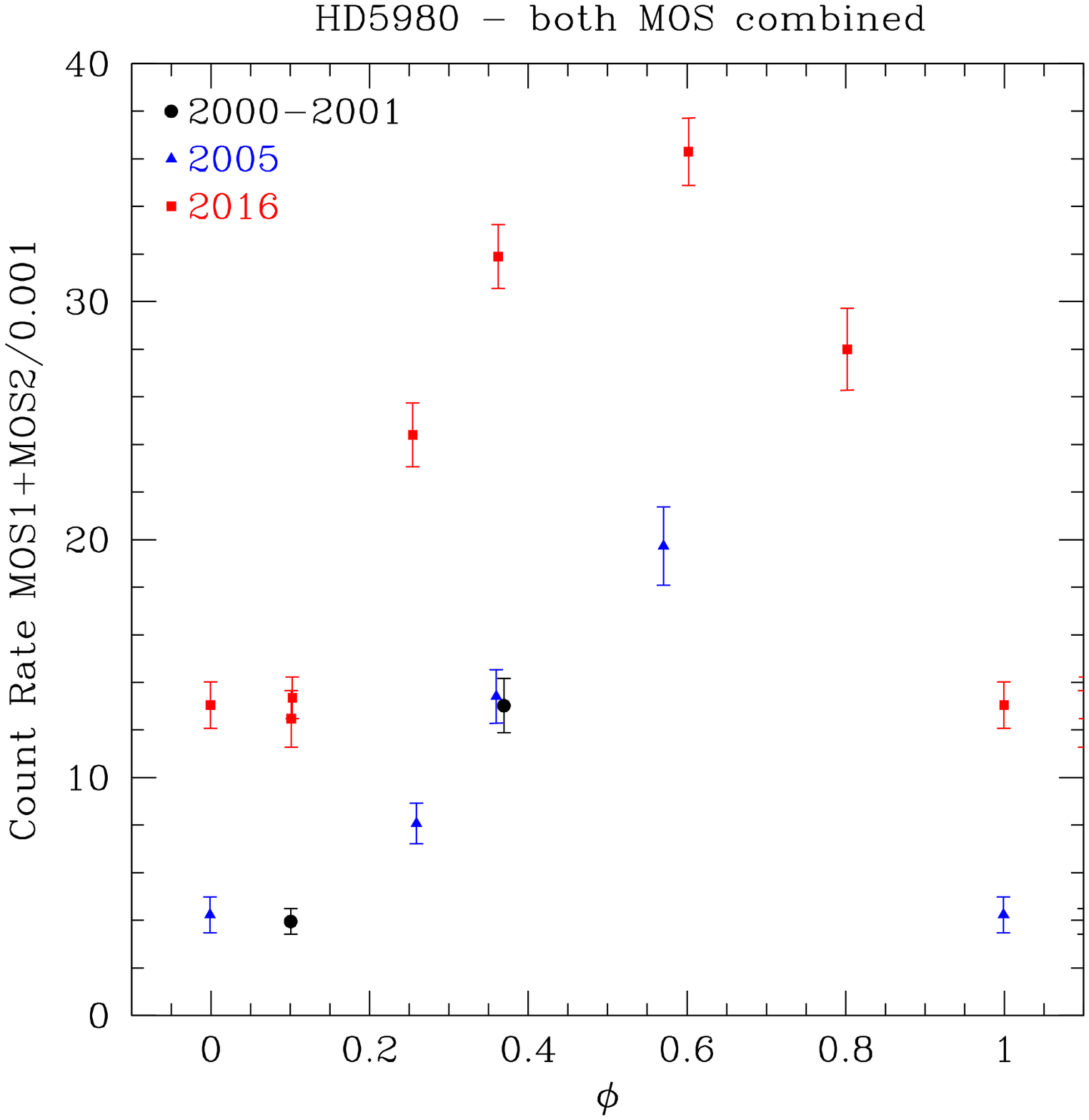}
\includegraphics[width=6.5cm]{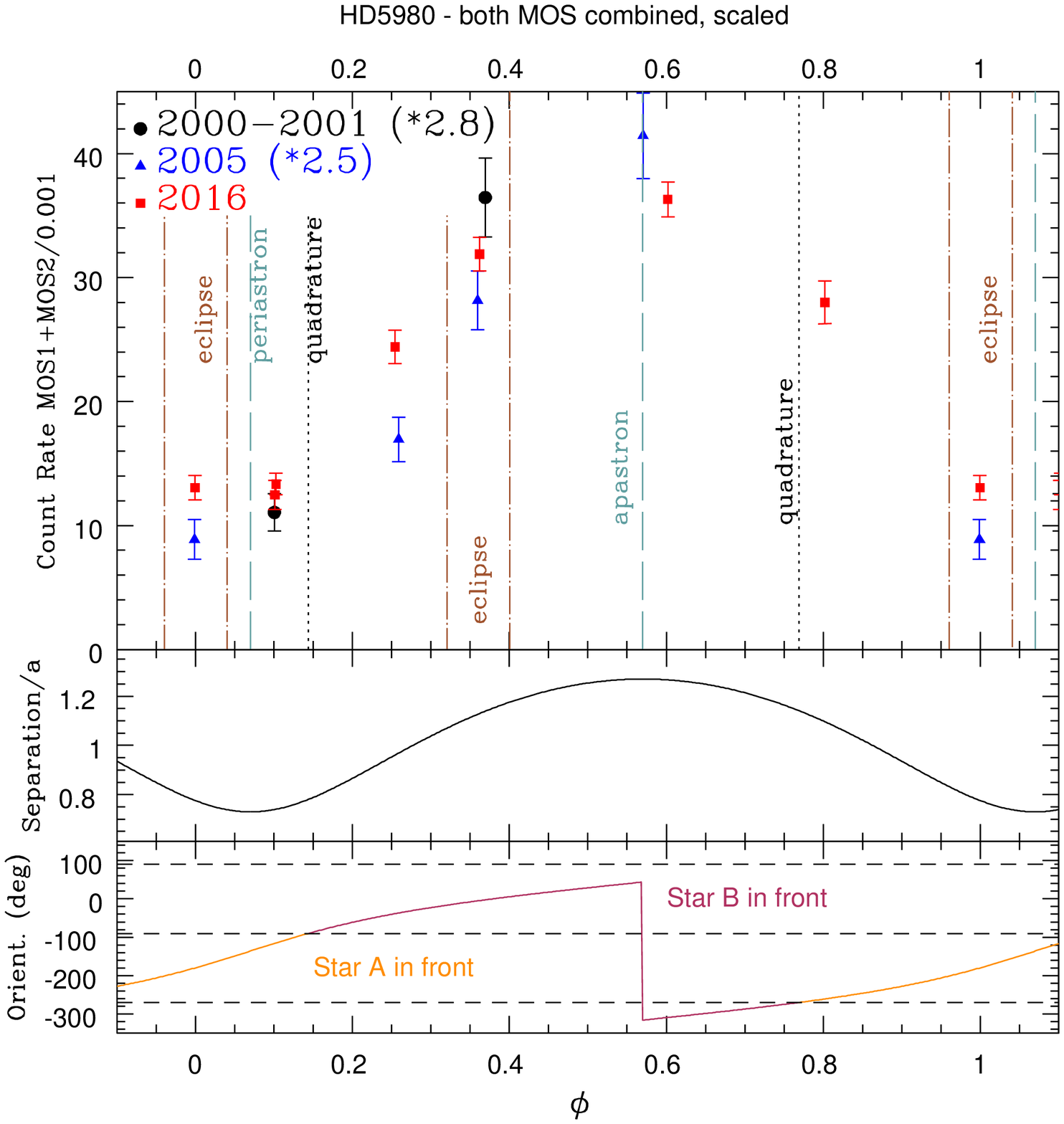}
\includegraphics[width=6cm]{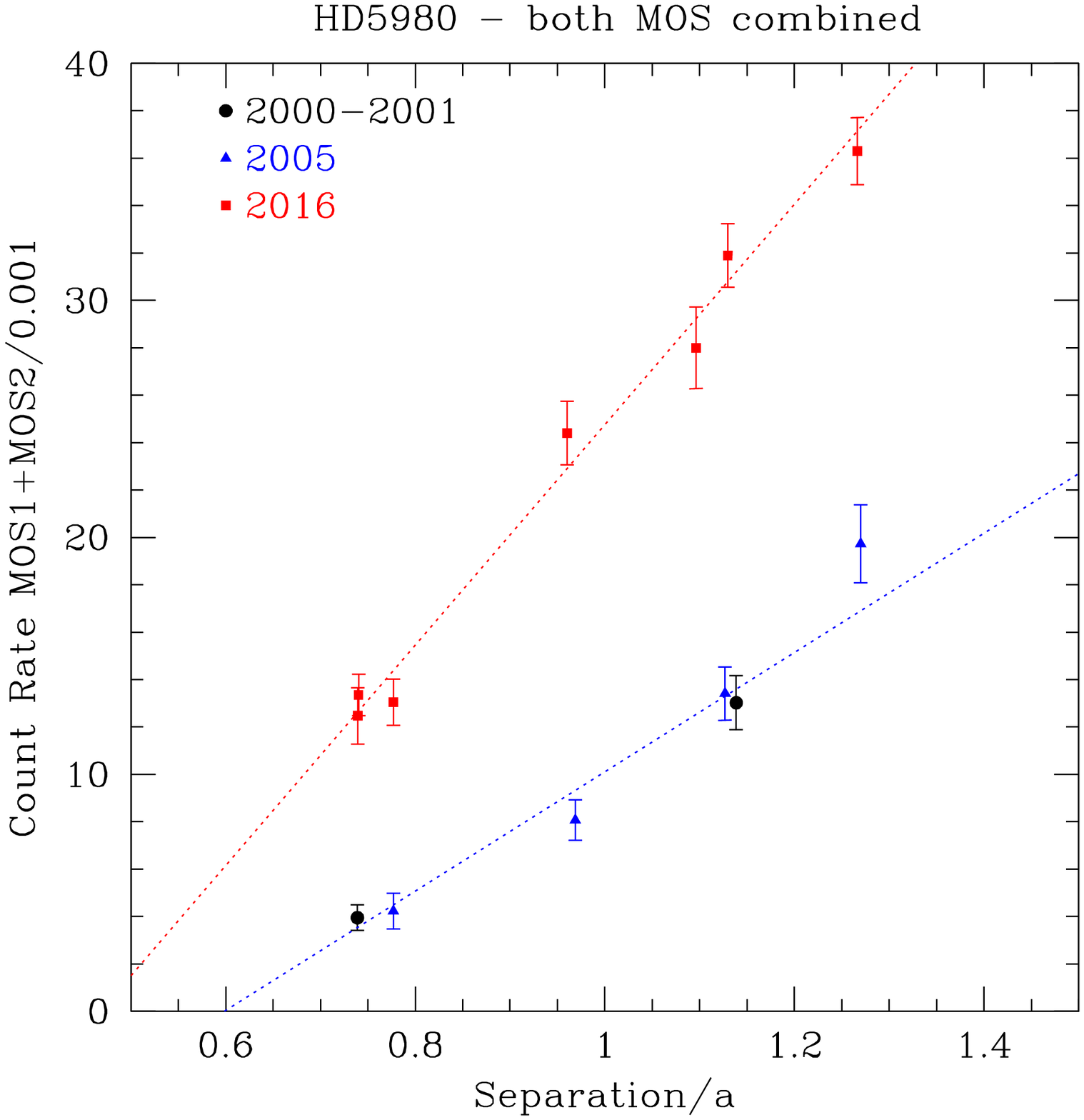}
\caption{{\it Left:} \xmm\ lightcurve of \hd\ in the 1.5--10.0\,keV energy band, for both MOS instruments combined. Data from 2000-2001 are shown as black circles, from 2005 as blue triangles, and from 2016 as red squares. Phases are computed using the ephemeris of \citet{ste97}. This panel shows the marked brightening of the X-ray emission in 2016. {\it Middle, from top to bottom:} \xmm\ lightcurve for both MOS combined using an arbitrary scaling factor to put all data points at similar levels (which demonstrates the remarkable similarity of the lightcurve shapes at different epochs); orbital separation (in units of the semimajor axis $a$); position angle (in $^{\circ}$) defined as 0$\degr$ when Star B is in front of Star A and 180$\degr$ in the opposite situation. The orbital solution is that of \citet{koe14}, i.e. $e=0.27$ and $\omega_{peri}=134^{\circ}$. The vertical lines in the top panel indicate the phases of the quadratures, periastron, apastron, and optical eclipses start/end. The horizontal dashed lines in the bottom panel indicate the quadratures. {\it Right:} Combined MOS count rates (without scaling) compared to the orbital separation (in units of the semimajor axis $a$), with their best-fit linear relations: a clear relationship is detected. }
\label{lc}
\end{figure*}

The first X-ray observations of \hd\ were performed with \ch\ and \xmm\ in 2000 and 2001 \citep{naz02,naz04}. These first exposures showed the source to be variable. Three additional observations obtained with \xmm\ in Fall 2005 confirmed the variability and further demonstrated its phased-locked nature since data taken on different dates but at the same phase appeared similar \citep [compare also black and blue symbols in Fig. \ref{lc}]{naz07}. In particular, the X-ray flux appeared larger around $\phi=0.36$, when Star B is in front of Star A, than at $\phi\sim0$. Such orbital variations indicate that the X-ray emission of \hd\ arises in a wind-wind collision. The absence of significant change in behaviour between data taken 5 years apart was particularly remarkable \citep{naz07}. 

The 2016 observations extend the X-ray coverage over the entire orbit (Table \ref{journal} and Fig. \ref{lc}). They confirm an increasing count rate from $\phi\sim0$ to $\phi=0.36$, but further show the rise in brightness to continue after the second conjunction up to the time of maximum stellar separation ($\phi\sim0.6$), with a decline thereafter. The tight correlation between the X-ray brightness of \hd\ and the orbital separation between Stars A and B is further illustrated in the right panel of Fig. \ref{lc}. The brightness variations appear to be linear in 2016, with an overall change by a factor of $\sim$3 between periastron and apastron. In previous years, the linear fit is slightly poorer and the relative amplitude of the variations appears larger (factor of $\sim$5), though the data point from Rev. 0970 is somewhat uncertain (see in Section 2.2). Note also that the shape of these variations does not seem to change much, as shown by the superposition of scaled lightcurves in the middle panel of Fig. \ref{lc}. 

This seems to indicate that orbital separation is the key parameter to explain variations, not orientation. Indeed, while the X-ray emission appears weak when Star A occults its companion ($\phi$=0), the interval of low flux lasts much longer than the optical eclipses and the expected occultation during the opposite eclipse (at $\phi$=0.36) is not observed (Fig. \ref{lc}, middle). On the other hand, one could rather expect an increase of X-ray flux only when Star B and its more tenuous wind is on the near side of the orbit ($\phi \sim 0.14 - 0.77$, with a peak at the conjunction $\phi=0.36$) but there is no such increase at those phases. This shows that neither stellar occultation nor large absorption changes in wind absorption produce the observed X-ray variations. 

However, the new observations also bring a surprise: the 2016 count rates are, on average, $2.57\pm0.12$ times larger than in previous epochs (Table \ref{journal} and left panel of Fig. \ref{lc}). We recall here that all \xmm\ observations were analyzed in the same way, to ensure homogeneity. Besides, the reported count rates correspond to the 1.5--10.\,keV energy band, to minimize the contamination by the SNR (see Sect. 2.2). Some remaining contamination may still exist, possibly slightly shifting upwards the count rates, but its Poissonian contribution is taken into account for the error calculation, hence the reported values should be correct, as best as possible considering the data properties and current knowledge of the instrument. The variations of \hd\ are thus no artifact.

\subsection{Spectral properties in X-rays}

The X-ray spectra were fitted within Xspec v12.9.0i, using as reference the solar abundances of \citet{asp09}. For \xmm, the three EPIC spectra (MOS1, MOS2, pn) were fitted simultaneously. In these fits, the emission from \hd\ was represented, as is adequate for massive stars, by an absorbed thermal emission model (``$apec$'') including up to two separate emission components. As mentioned in Sect. 2.2 above, two different backgrounds were considered, the immediate surroundings of \hd\ (i.e. the SNR IKT\,18) and an external background (in which case the SNR contamination was fixed to its best-fit in \ch\ data). For the latter case, a single thermal component was sufficient to achieve a good fit, while two components were generally required for the former case. However, to facilitate comparison and evaluate the impact of the background choice on the results, we also performed a fitting with a single component for the SNR background choice. In addition, the metallicity of the thermal emission components was fixed to the best fit values found by Hillier et al. (in prep) for Star A: mass fractions X=0.2011, Y=0.7962, Z(C)=3.05$\times10^{-5}$, Z(N)=2.20$\times10^{-3}$, Z(O)=1.50$\times10^{-6}$, and Z(others)=0.20$\times$solar, corresponding to abundances in number, with respect to Hydrogen and with respect to solar, of 11.6 for He and N, 0.05 for C, 0.001 for O, and 0.734 for other elements. Two possibilities were then envisaged for the abundances of the absorption component: solar abundances (i.e. $phabs$, all absorption occurring in the Milky Way) and abundances fixed to the same values as for the emission component(s) (i.e. $vphabs$, all absorption being circumstellar). 

\begin{figure*}
\includegraphics[width=6cm]{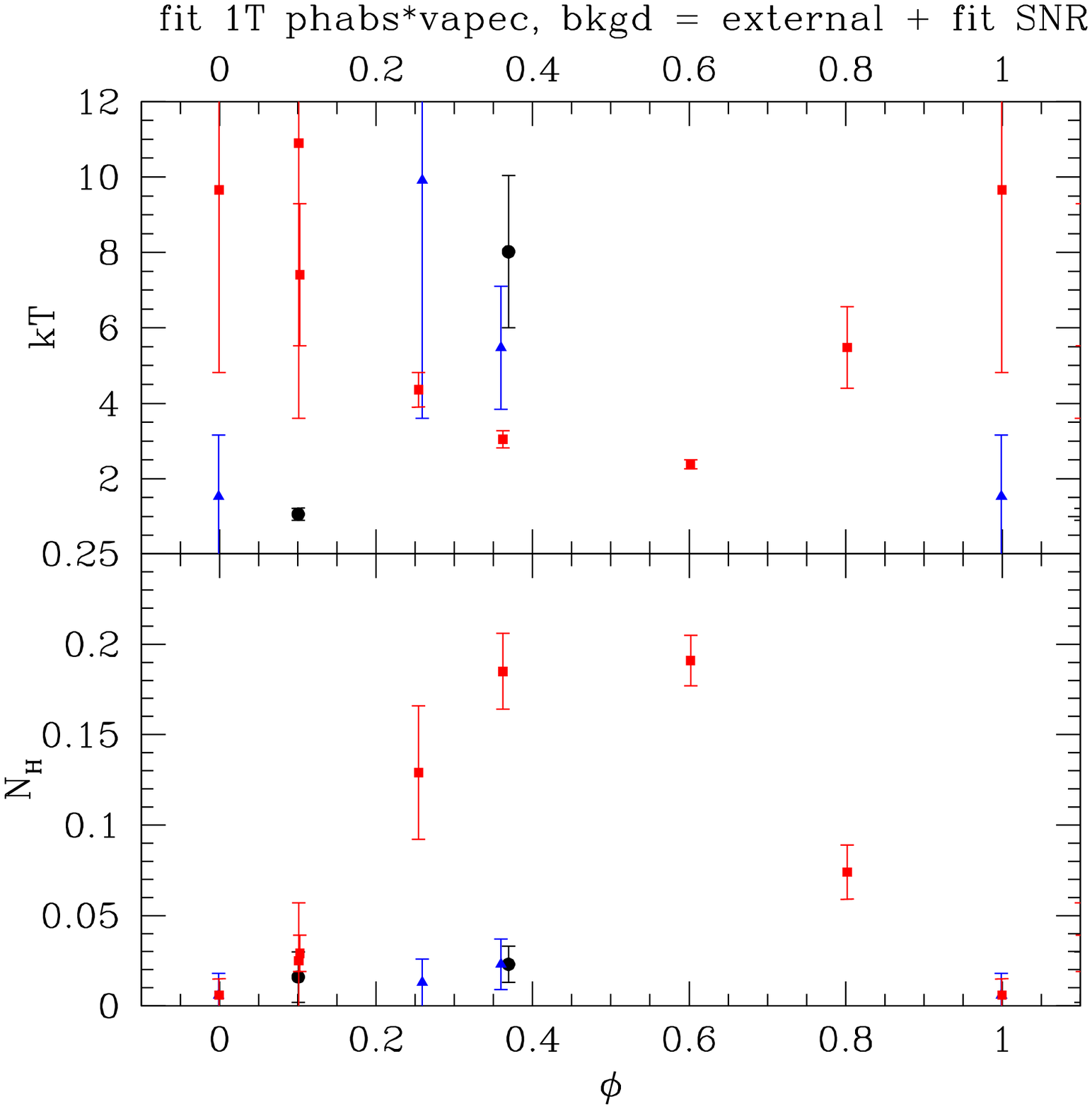}
\includegraphics[width=6cm]{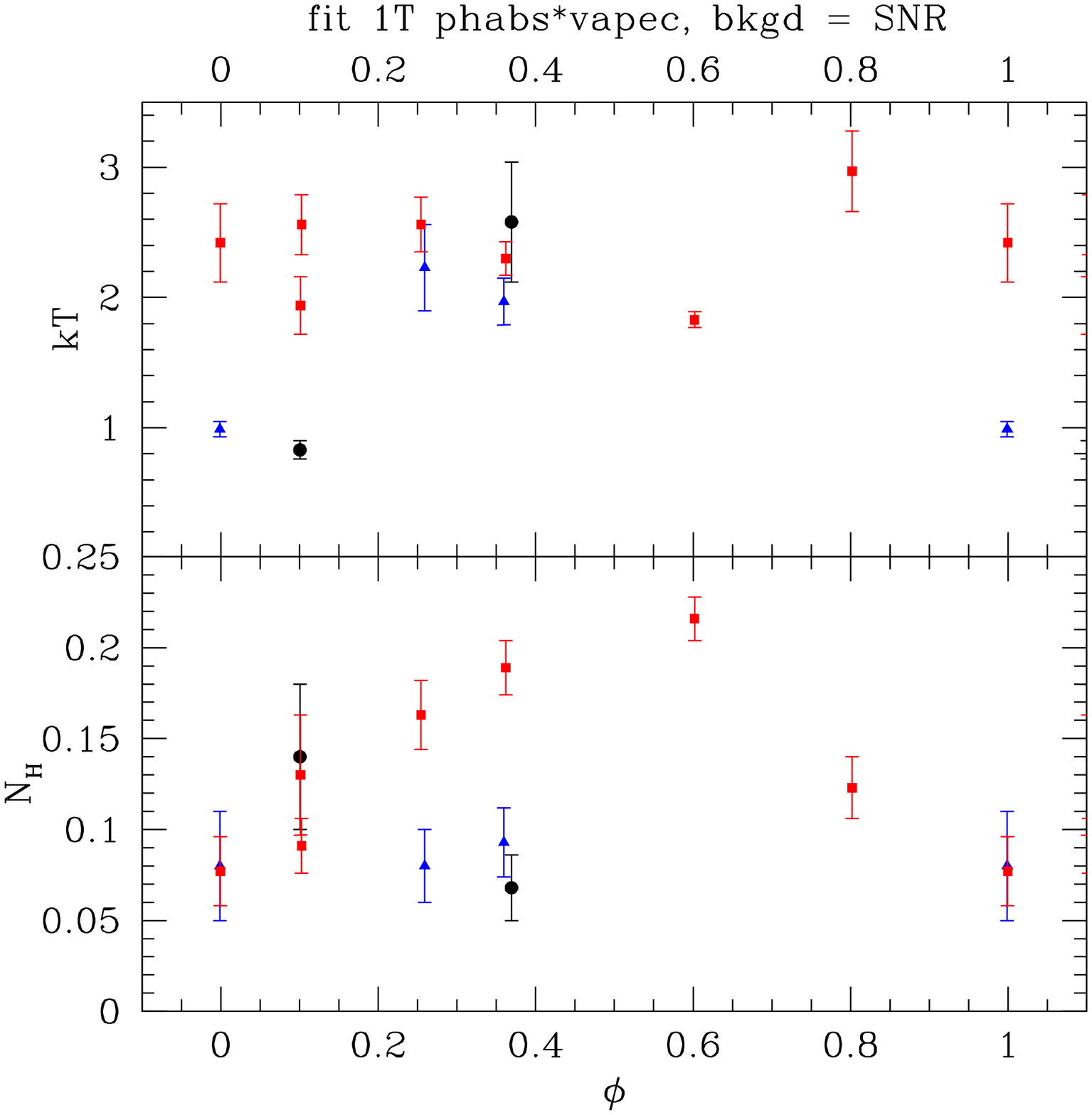}
\includegraphics[width=6cm]{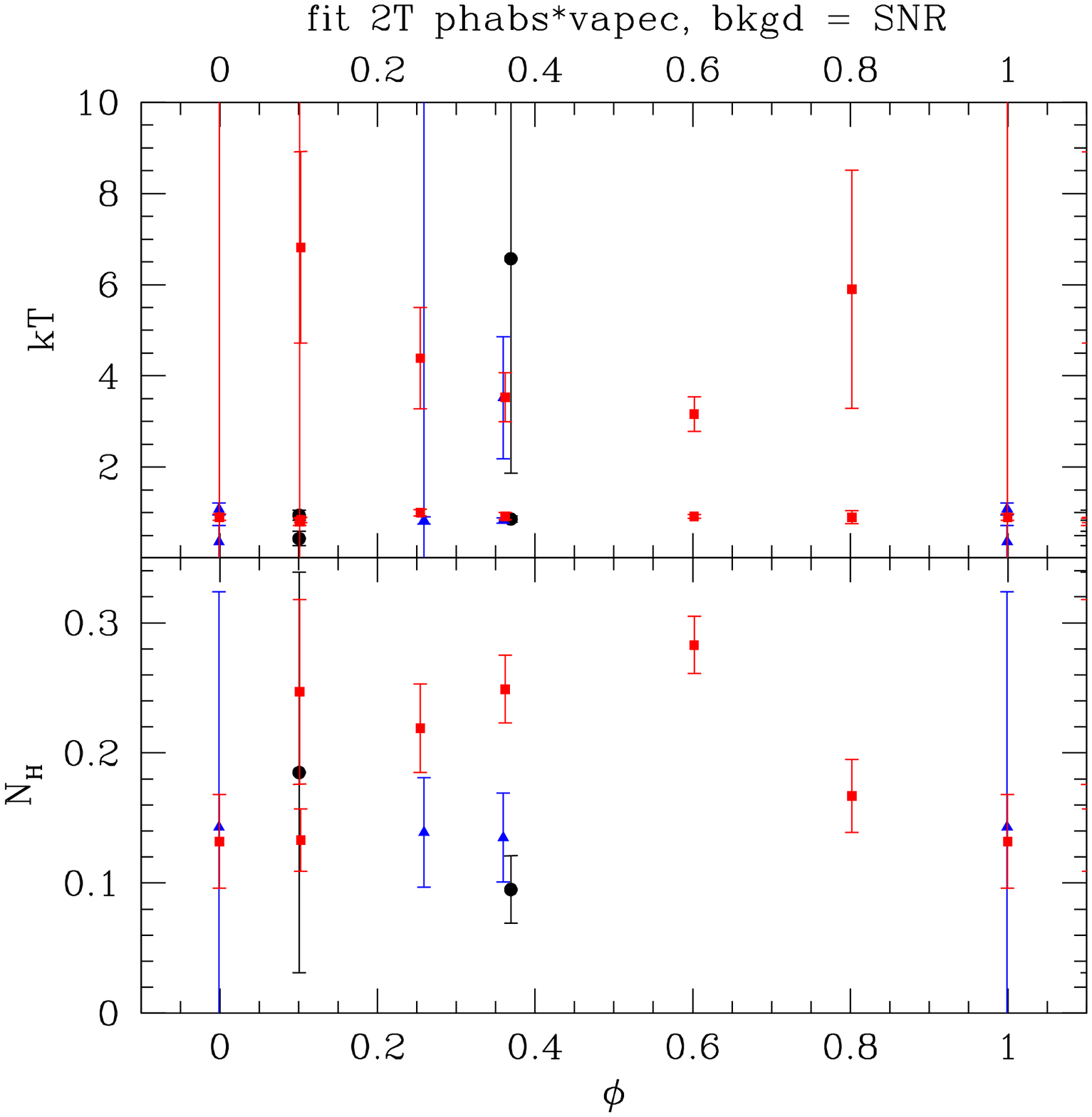}
\caption{Evolution with phase of the spectral parameters derived for \xmm\ spectra considering solar absorption (see Table \ref{fit}): left, 1T results when choosing an external background and a SNR contamination fixed to \ch\ best fit; middle and right, 1T and 2T results, respectively, for the choice of the surrounding SNR as background.  }  
\label{fitparam}
\end{figure*}

The results of the fits for the different methods (different absorptions, 1 or 2 thermal components, external or SNR background) are presented in Table \ref{fit} and Fig. \ref{fitparam}. We note that the choice of the background or the choice of absorption abundances does not change the derived phase-locked trends - it only shifts the parameters' averages. It is also immediately obvious that the flux variations derived from the count rates are confirmed. However, results for \xmm\ spectra taken in Revs. 0357 and 1100 differ from other ones. In fact, at these phases, the flux from \hd\ is minimal, hence its emission is more difficult to disentangle from the SNR contamination. Clearly, some residual contamination from the SNR remains, whatever the method used for background subtraction, leading to (unrealistic) lower temperatures at these phases. Another evidence of residual contamination lies in the fluxes: those derived from \ch\ data are systematically lower than those calculated from \xmm\ spectra (for a similar phase and observing year). While this affects the absolute values of the spectral parameters, it does not change their variations, as SNR emission remains constant over such a short time span.

\begin{figure*}
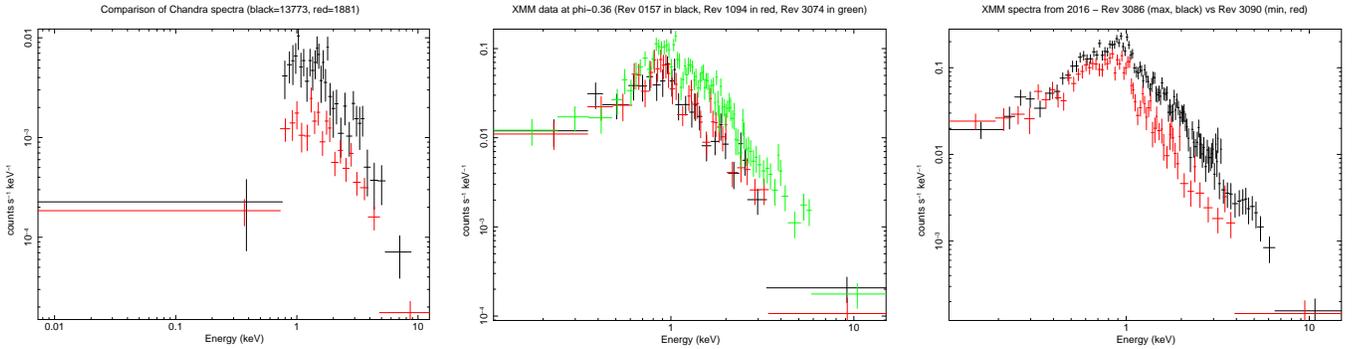

\includegraphics[height=6cm, angle=270]{newcompachandra.ps}
\includegraphics[height=6cm, angle=270]{newcompaspecxmmphi.ps}
\includegraphics[height=6cm, angle=270]{newcompaspecxmm2016b.ps}
\caption{Spectra from \hd, corrected for the (SNR) background, as recorded by \ch\ (left) and EPIC-pn onboard \xmm\ (middle panel for spectra taken at a similar phase but in different years; right panel for spectra taken at minimum and maximum brightness levels in 2016). The hardening of the spectra in recent years is obvious in the middle panel.}  
\label{compaspec}
\end{figure*}

\begin{table*}
\centering
\caption{Results of the spectral fits using models of the type $(v)phabs\times \sum vapec_i$ . Fluxes are provided in the 0.5--10.0\,keV energy band, the usual range used for massive stars. Limits of 1$\sigma$ (or 68\%) confidence intervals are provided within squared brackets but errors on fluxes should be taken with caution, as uncertainties linked to cross-calibration and remaining SNR contamination subsist for this parameter. Note that, formally, spectra from Revs. 0357 and 1100 do not require two thermal components to be well fitted (those fits are here only provided for homogeneity and completeness). }
\label{fit}
\begin{tabular}{lcccccccc}
\hline\hline
ID/Rev  & $\phi$ & $N_{\rm H}$ & $kT_1$ & $norm_1$ & $kT_2$ & $norm_2$ & $\chi^2_{\nu}$ (dof) & $F^{\rm obs}_{\rm X}$ \\
    &        & (10$^{22}$\,cm$^{-2}$) & (keV) & ($10^{-5}$\,cm$^{-5}$) & (keV) & ($10^{-5}$\,cm$^{-5}$) & & ($10^{-14}$\,erg\,cm$^{-2}$\,s$^{-1}$)\\
\hline
\multicolumn{9}{l}{1T, solar absorption, using an external background and a fixed model for the SNR contamination}\\
0157 & 0.369 &0.023 [0.015-0.033] &8.02 [6.11-11.0] &1.89 [1.75-2.02] & & & 1.29(130) &10.9 [10.1-11.2]\\
0357 & 0.100 &0.016 [0.004-0.030] &1.06 [0.94-1.22] &0.69 [0.58-0.79] & & & 1.94(122) &1.80 [1.76-1.84]\\
1093 & 0.259 &0.013 [0.003-0.026] &9.91 [6.85-16.2] &1.35 [1.25-1.46] & & & 1.23(119) &8.15 [7.35-8.40]\\
1094 & 0.360 &0.023 [0.013-0.037] &5.47 [4.42-7.10] &1.90 [1.76-2.05] & & & 1.59(130) &9.80 [9.15-10.2]\\
1100 & 0.999 &0.006 [0.000-0.018] &1.53 [1.08-3.16] &0.77 [0.63-0.91] & & & 1.52(88)  &2.20 [2.12-2.28]\\
3073 & 0.254 &0.129 [0.109-0.156] &4.36 [3.90-4.86] &4.61 [4.40-4.86] & & & 1.20(189) &19.9 [19.3-20.4]\\
3074 & 0.362 &0.185 [0.165-0.206] &3.05 [2.83-3.28] &7.28 [6.97-7.60] & & & 1.40(231) &25.1 [24.5-25.6]\\
3086 & 0.602 &0.191 [0.178-0.205] &2.38 [2.27-2.50] &10.3 [9.97-10.6] & & & 1.37(255) &30.1 [29.5-30.7]\\
3088 & 0.802 &0.074 [0.060-0.091] &5.48 [4.77-6.56] &4.73 [4.51-4.97] & & & 1.13(175) &23.5 [22.1-24.5]\\
3090 & 0.999 &0.006 [0.000-0.015] &9.66 [7.14-14.5] &1.60 [1.50-1.70] & & & 1.32(147) &9.67 [9.03-9.99]\\
3091 & 0.101 &0.025 [0.009-0.057] &10.9 [7.25-18.2] &1.51 [1.44-1.64] & & & 1.28(113) &9.15 [7.63-9.49]\\
3110 & 0.102 &0.029 [0.021-0.039] &7.41 [6.30-9.29] &2.06 [1.96-2.16] & & & 1.61(180) &11.6 [11.1-12.0]\\
\hline
\multicolumn{9}{l}{1T, solar absorption, using the surrounding SNR as background}\\
1881  &0.266 &0.19 [0.11-0.30] &7.62 [5.44-12.1] &0.70 [0.63-0.79] &&&1.39(17)  &3.65 [2.89-3.89] \\
13773 &0.823 &0.22 [0.12-0.33] &4.51 [3.48-6.70] &5.39 [4.72-6.06] &&&1.01(29)  &22.5 [18.2-24.4] \\
0157  &0.369 &0.07 [0.05-0.09] &2.58 [2.27-3.03] &3.35 [3.10-3.58] &&&1.67(54)  &11.5 [10.9-12.2] \\
0357  &0.100 &0.14 [0.10-0.18] &0.83 [0.78-0.90] &2.76 [2.48-3.09] &&&1.52(39)  &5.54 [5.23-5.76] \\
1093  &0.259 &0.08 [0.06-0.10] &2.23 [1.90-2.56] &2.77 [2.54-3.00] &&&2.17(44)  &8.65 [7.98-9.15] \\
1094  &0.360 &0.09 [0.08-0.11] &1.97 [1.80-2.15] &3.79 [3.56-4.01] &&&1.50(60)  &10.8 [10.2-11.4] \\
1100  &0.999 &0.08 [0.06-0.11] &0.99 [0.93-1.06] &2.52 [2.28-2.77] &&&0.61(19)  &5.87 [5.45-6.25] \\
3073  &0.254 &0.16 [0.15-0.18] &2.56 [2.39-2.78] &6.49 [6.19-6.79] &&&1.36(124) &20.4 [19.6-21.1] \\
3074  &0.362 &0.19 [0.18-0.20] &2.30 [2.18-2.43] &9.23 [8.91-9.55] &&&1.41(175) &26.4 [25.6-27.2] \\
3086  &0.602 &0.22 [0.21-0.23] &1.83 [1.77-1.89] &12.7 [12.3-13.0] &&&1.63(202) &30.4 [29.8-31.2] \\
3088  &0.802 &0.12 [0.11-0.14] &2.97 [2.66-3.27] &6.44 [6.13-6.75] &&&1.40(112) &22.9 [21.8-24.0] \\
3090  &0.999 &0.08 [0.06-0.10] &2.42 [2.12-2.72] &2.97 [2.77-3.17] &&&2.22(65)  &9.77 [9.18-10.4] \\
3091  &0.101 &0.13 [0.10-0.16] &1.94 [1.72-2.16] &3.40 [3.13-3.67] &&&2.50(41)  &9.26 [8.59-9.82] \\
3110  &0.102 &0.09 [0.07-0.11] &2.56 [2.33-2.79] &3.61 [3.43-3.79] &&&2.03(103) &12.1 [11.6-12.6] \\
\multicolumn{9}{l}{2T, solar absorption, using the surrounding SNR as background}\\
0157 &0.369 &0.10 [0.08-0.12] &0.86 [0.80-0.93] &2.04 [1.68-2.44] &6.57 [4.54-11.3] &1.72 [1.44-2.01] &1.07(52)  &13.5 [11.6-13.9] \\
0357 &0.100 &0.19 [0.13-0.34] &0.43 [0.30-0.59] &1.74 [0.68-7.40] &0.94 [0.86-1.06] &2.08 [1.61-2.58] &1.53(37)  &5.67 [2.67-5.44] \\
1093 &0.259 &0.14 [0.11-0.18] &0.83 [0.76-0.89] &2.40 [2.01-2.87] &10.8 [5.95-48.7] &1.06 [0.91-1.27] &1.14(42)  &10.9 [4.26-11.3] \\
1094 &0.360 &0.14 [0.11-0.17] &0.83 [0.79-0.89] &2.39 [2.00-2.88] &3.52 [2.97-4.86] &1.90 [1.55-2.20] &0.87(58)  &12.2 [10.8-12.8] \\
1100 &0.999 &0.14 [0.07-0.32] &0.36 [0.24-0.72] &1.82 [0.32-28.0] &1.08 [0.97-1.21] &2.12 [1.74-2.69] &0.58(17)  &6.05 [3.70-5.90] \\
3073 &0.254 &0.22 [0.19-0.25] &1.00 [0.94-1.07] &3.49 [2.92-4.15] &4.39 [3.76-5.50] &3.79 [3.27-4.26] &1.03(122] &22.0 [20.5-22.6] \\
3074 &0.362 &0.25 [0.23-0.28] &0.92 [0.84-0.98] &4.51 [3.84-5.22] &3.53 [3.15-4.07] &5.69 [5.12-6.32] &1.02(173) &28.3 [26.8-29.0] \\
3086 &0.602 &0.28 [0.26-0.31] &0.92 [0.88-0.96] &7.69 [6.84-8.58] &3.16 [2.81-3.54] &6.30 [5.56-7.09] &0.99(200) &33.2 [32.2-33.8] \\
3088 &0.802 &0.17 [0.14-0.20] &0.90 [0.77-0.98] &3.19 [2.56-3.86] &5.90 [4.50-8.51] &4.03 [3.51-4.66] &1.06(110) &25.7 [23.2-26.5] \\
3090 &0.999 &0.13 [0.10-0.17] &0.90 [0.84-0.97] &2.36 [1.99-2.75] &11.6 [6.48-35.1] &1.26 [1.09-1.48] &1.44(63)  &12.2 [4.59-12.7] \\
3091 &0.101 &0.25 [0.19-0.32] &0.80 [0.72-0.88] &3.57 [2.89-4.51] &16.0 [7.13-64.0] &1.17 [0.99-1.37] &1.36(39)  &12.3 [5.00-12.8] \\
3110 &0.102 &0.13 [0.11-0.16] &0.84 [0.79-0.91] &2.32 [2.03-2.62] &6.82 [5.21-8.92] &1.87 [1.69-2.09] &1.32(101) &14.4 [13.3-14.8] \\
\multicolumn{9}{l}{2T, non-solar absorption, using the surrounding SNR as background}\\
0157 &0.369 &0.014 [0.011-0.018] &0.86 [0.81-0.96] &1.69 [1.40-2.09] &6.02 [4.44-10.9] &1.79 [1.46-2.06] &1.06(52)  &13.5 [11.5-14.1] \\
0357 &0.100 &0.020 [0.015-0.027] &0.83 [0.78-0.89] &2.24 [2.04-2.46] &68.4 [2.91-64.0] &0.25 [0.17-0.38] &1.52(37)  &6.49 [4.67-15.8] \\
1093 &0.259 &0.021 [0.016-0.029] &0.85 [0.78-0.91] &1.96 [1.68-2.27] &9.63 [5.58-27.9] &1.11 [0.95-1.31] &1.09(42)  &11.0 [7.66-11.4] \\
1094 &0.360 &0.020 [0.016-0.025] &0.86 [0.78-0.90] &1.85 [1.57-2.20] &3.52 [3.00-4.70] &1.93 [1.62-2.20] &0.87(58)  &12.3 [10.9-12.7] \\
1100 &0.999 &0.019 [0.010-0.072] &0.36 [0.19-0.70] &1.16 [0.20-35.9] &1.11 [1.01-1.25] &2.01 [1.41-2.42] &0.56(17)  &6.09 [4.51-6.21] \\
3073 &0.254 &0.036 [0.030-0.042] &1.06 [0.99-1.12] &2.44 [2.00-2.92] &4.44 [3.85-5.45] &3.82 [3.37-4.23] &1.03(122) &22.2 [20.9-22.9] \\
3074 &0.362 &0.043 [0.038-0.048] &0.97 [0.90-1.03] &2.95 [2.45-3.47] &3.60 [3.24-4.09] &5.71 [5.21-6.22] &1.03(173) &28.6 [27.2-29.4] \\
3086 &0.602 &0.049 [0.045-0.053] &0.97 [0.92-1.01] &4.96 [4.35-5.59] &3.10 [2.80-3.44] &6.61 [5.95-7.27] &0.95(200) &33.5 [32.5-34.2] \\
3088 &0.802 &0.026 [0.022-0.031] &0.94 [0.82-1.02] &2.39 [1.87-2.93] &5.90 [4.68-8.25] &4.04 [3.57-4.54] &1.02(110) &25.9 [23.6-26.8] \\
3090 &0.999 &0.021 [0.016-0.027] &0.93 [0.85-1.00] &2.00 [1.69-2.31] &10.8 [6.45-29.6] &1.27 [1.11-1.49] &1.43(63)  &12.3 [4.64-12.9] \\
3091 &0.101 &0.045 [0.032-0.062] &0.82 [0.74-0.89] &2.71 [2.24-3.28] &12.2 [6.36-64.0] &1.23 [1.05-1.44] &1.31(39)  &12.4 [4.65-12.7] \\
3110 &0.102 &0.020 [0.017-0.024] &0.87 [0.81-0.94] &1.86 [1.62-2.13] &6.56 [5.09-8.79] &1.91 [1.70-2.13] &1.27(101) &14.3 [13.4-14.9] \\
\hline
\end{tabular}
\end{table*}

Regarding spectral parameters derived in 2016 (the time of our best coverage of the orbit), we find that the absorption column increases towards apastron while the temperature decreases (only that of the hottest thermal component if two were used, see Table \ref{fit} and Fig. \ref{fitparam}). These results are however reminiscent of the usual absorption-temperature degeneracy: spectra can be reproduced by a low-absorption, high-temperature model as well as by a high-absorption, low-temperature model \citep[e.g.][]{naz12b}. To remove this degeneracy, we fixed the temperatures and performed the fits again: the absorption still appears larger at apastron. We also did a trial in which absorption was fixed to its mean value: the temperature still appears smaller at apastron. This shows that our result goes beyond the trade-off problem: there are truly an increase of absorption and a decrease of temperature at apastron.

Since observations span over 15 years, we may also assess whether there have been any changes of the spectral parameters with time, by comparing spectra taken at similar orbital phases in different orbital cycles. For example, the first \ch\ dataset and two \xmm\ exposures (Revs. 1093 and 3073) were taken at $\phi\sim0.25$, data in Revs. 0157, 1094, and 3074 were taken at an eclipse of Star A by Star B ($\phi\sim0.36$) while the second \ch\ observation was taken at the same time as \xmm\ data of Rev. 3088. We need however to keep in mind that (1) there may be remaining cross-calibration problems and some SNR contamination in \xmm\ spectra, which leaves some uncertainty for any \xmm\ vs \ch\ comparison and (2) the lower luminosities of \hd\ observed in 2000--2005 renders the parameter determination more difficult at those times, hence they are less precise. 

For \xmm\ spectra, there is a clear hardening of the emission of \hd\ in 2016 compared to that recorded in previous years (see also middle panel of Fig. \ref{compaspec}). This hardening does not seem to be linked to changing temperatures (since they appear quite similar whatever the year considered) but rather to an increase in absorbing columns with time (Table \ref{fit}, Fig. \ref{fitparam}). However, the larger noise in older data complicates the comparison so our interpretation of the origin of the observed hardening should be viewed with some caution. In contrast, the fits of both \ch\ spectra, which have minimal contamination by the SNR, show only a slight increase in absorption and a slight decrease in temperature, formally not significant as they remain within 1$\sigma$. For them, there is thus mostly a flux variation (see also left panel of Fig. \ref{compaspec}). It is however important to note that these exposures were taken at different phases ($\phi$=0.27 and 0.82) and the spectra are unfortunately noisier than the \xmm\ ones, rendering detailed comparisons difficult.

\subsection{Discussion}

\begin{table}
\centering
\footnotesize
\caption{Stellar parameters derived for Star A from the analysis with CMFGEN of HST data taken in 2000, 2002, 2009 \citep{geo11} and 2014 (Hillier et al., in prep). Mass-loss rate values reported here consider a filling factor $f$ of 0.1, typical of WR stars. Simple interpolations (in italics) were calculated for 2001 and 2005. The typical errors reported in the second column correspond to formal uncertainties on the match between the CMFGEN model and the observations; they do not take into account the systematic uncertainties in the underlying assumptions inherent to all atmosphere models currently available, the non totality of the eclipses, or emission arising from the wind-wind collision.}
\label{starA}
\begin{tabular}{lccccccc}
\hline\hline
Parameter & $\sigma$ & 2000 & {\it 2001} & 2002 & {\it 2005} & 2009 & 2014 \\
\hline
$\log(L_{\rm BOL}/L_{\odot})$                & 0.1 & 6.30 & {\it 6.35} & 6.39 & {\it 6.39} & 6.39 & 6.23 \\
$T_{\rm eff}$ (kK)                          & 3.0 & 37.3 & {\it 38.7} & 40.0 & {\it 41.5} & 43.0 & 43.0\\
$R_*$ (R$_{\odot}$)                         & 10\% & 34   & {\it 33}   & 32   & {\it 30}   & 28   & 21.6\\
$\dot M$ ($10^{-5}$\,M$_{\odot}$\,yr$^{-1}$) & 10\% & 11.1 & {\it 9.5}  & 7.9  & {\it 7.6}  & 7.3  & 5.0 \\
$v_{\infty}$ (km\,s$^{-1}$)                  & 200 & 2000 & {\it 2100} & 2200 & {\it 2320} & 2440 & 3000\\
\hline
\end{tabular}
\end{table}

The previous sections have helped us derive the observational picture; we now summarize the main observational results and propose our interpretation. 

\begin{figure}
\includegraphics[width=8cm]{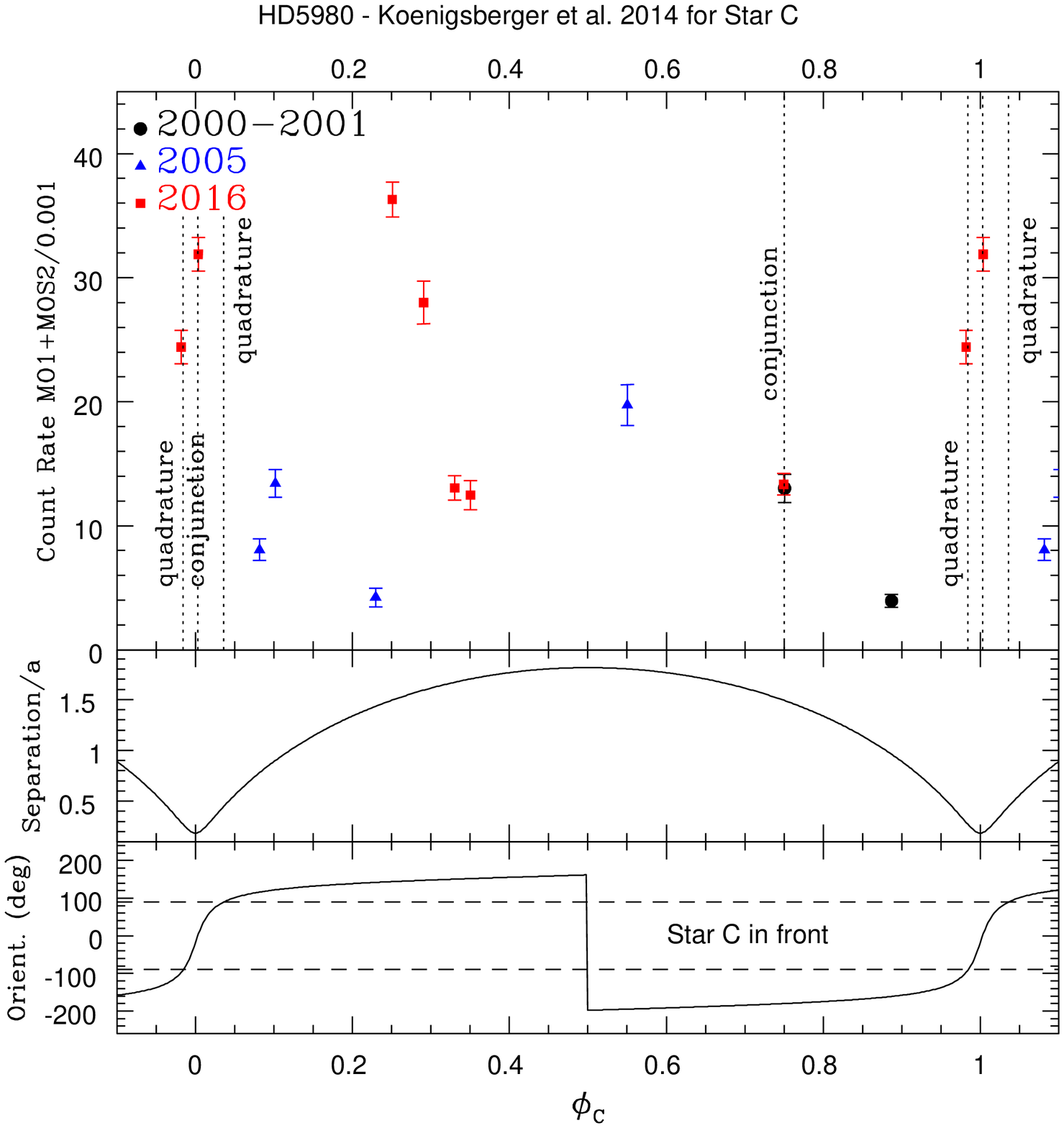}
\caption{Same as Fig. \ref{lc} but using the ephemeris relative to Star C \citep[see Table 7 in][ and 6th column of Table \ref{journal} above]{koe14}. There is no smooth, coherent behavior with phase.}  
\label{starc}
\end{figure}

First and foremost, we discover an overall flux increase over the last decade, by a factor of $\sim$2.5, while data taken between 2000 and 2005 showed no significant differences in this respect. This brightening is accompanied by a slight hardening of the X-ray emission, possibly linked to an increased absorption. Such a brightening is totally unprecedented in colliding wind binaries. Repeatability rather is the rule \citep[see e.g. the case of WR21a][]{gos16} - only $\eta$\,Car presents changes from cycle-to-cycle, but only over a very small phase interval \citep[around periastron, see e.g.][]{ham14}. Since it is unprecedented, this secular change may cast doubt that the recorded X-rays truly are associated with the wind-wind collision in the A+B pair, but alternative explanations can all be ruled out. Indeed, the observed X-ray increase cannot be produced by the collision of the now fast wind from Star A catching up to and colliding with the slower material ejected in the eruption, since this would lead to a brightening by a constant amount at all phases (i.e. a simple shift of the lightcurve towards larger luminosities) at odds with observations showing a brightening by a similar factor at all phases (Fig. \ref{lc}). The variation cannot be associated with Star C either: when the observed variation are phased with the Star C ephemeris \citep[see also column $\phi_C$ in Table \ref{journal}]{koe14}, there is no coherent behaviour with phase (Fig. \ref{starc}), and incompatibilities are even present at $\phi_C\sim0.2-0.4$. Similarly, a combination of two collisions, one in the A+B pair and one in the C+? system, can also be ruled out. Indeed, Table \ref{journal} shows that some observations were taken with similar phases in both orbits: if the variations were due to a combination of two periodic variability schemes, data taken at similar phase pairs would be similar, but this is not the case (e.g. data taken in Revs. 1093 and 3073 widely differ in flux!). Thus the changing wind-wind collision in the A+B pair must be responsible for the observed recent changes in the X-ray emission. In this context, it is important to note that X-ray spectra from 2000 and 2005, obtained years apart but at the same orbital phase, appeared similar in brightness and spectral shape to each other, demonstrating a clear phase-locked repeatability, and indicating that the large-scale increase in X-ray brightness we observed in 2016 occurred rather suddenly after 2005.

Second, we confirm the presence of phase-locked flux variations detected earlier and refine their properties. In particular, we find a tight correlation between the X-ray brightness and the orbital separation. This result clearly indicates that the collision is not adiabatic, since a maximum at periastron is expected in such cases (see e.g. the cases of WR25 \citealt{gosth,pan14}, or Cyg\,OB2\,\#9 \citealt{naz12c}). A similar behaviour was however recorded in several {\it radiative} wind-wind collisions (e.g. Cyg\,OB2\,\#8A - see Fig. 3 of \citealt{caz14}, HD\,152218 and HD\,152248 - see Fig. 3 in \citealt{rau16}, HD\,166734 - see Fig. 5 in \citealt{naz17}) though the luminosity-separation trend is here much more linear and the hysteresis around it of a much smaller amplitude (see right panel of Fig. \ref{lc}). 

Another result to point out is the absence of variations related to the orientation of the system. Not only is there no evidence of eclipses \citep[unlike the case of V444\,Cygni, see][but similar to the case of WR\,20a, see \citealt{naz08}, and CQ\,Cep, see \citealt{ski15}]{lom15}, but there is also no evidence of the usual absorption effects. Indeed, as far as we can tell, there is no significant increase of absorption at or close to conjunctions (i.e. at eclipse phases) nor at periastron, contrary to what is observed e.g. for the radiative collisions in V444\,Cygni \citep{lom15} or WR21a \citep{gos16}. Furthermore, we fail to detect a significantly lower absorption, or larger flux, when the secondary star (Star B) and its more tenuous wind are in front of the system, though (1) this should not occur in the new monitoring as both stars now have similar wind parameters (see below) and (2) it is true that such effects mostly affect soft X-rays (see e.g. $\gamma^2$\,Vel, \citealt{wil95}) which are somewhat uncertain in our case as they are most affected by any remaining SNR contamination. These facts, combined to the generally low absorbing columns found in fits, suggest that the X-ray emission zone is large compared to the stellar bodies and the innermost, opaque, parts of their stellar winds. It is however puzzling to find a larger absorption at apastron: considering that stars are then more distant, and far from conjunctions, the wind density along the line-of-sight should be smaller, not larger, underlining the peculiarity of this result.

\begin{figure*}
\includegraphics[width=9cm]{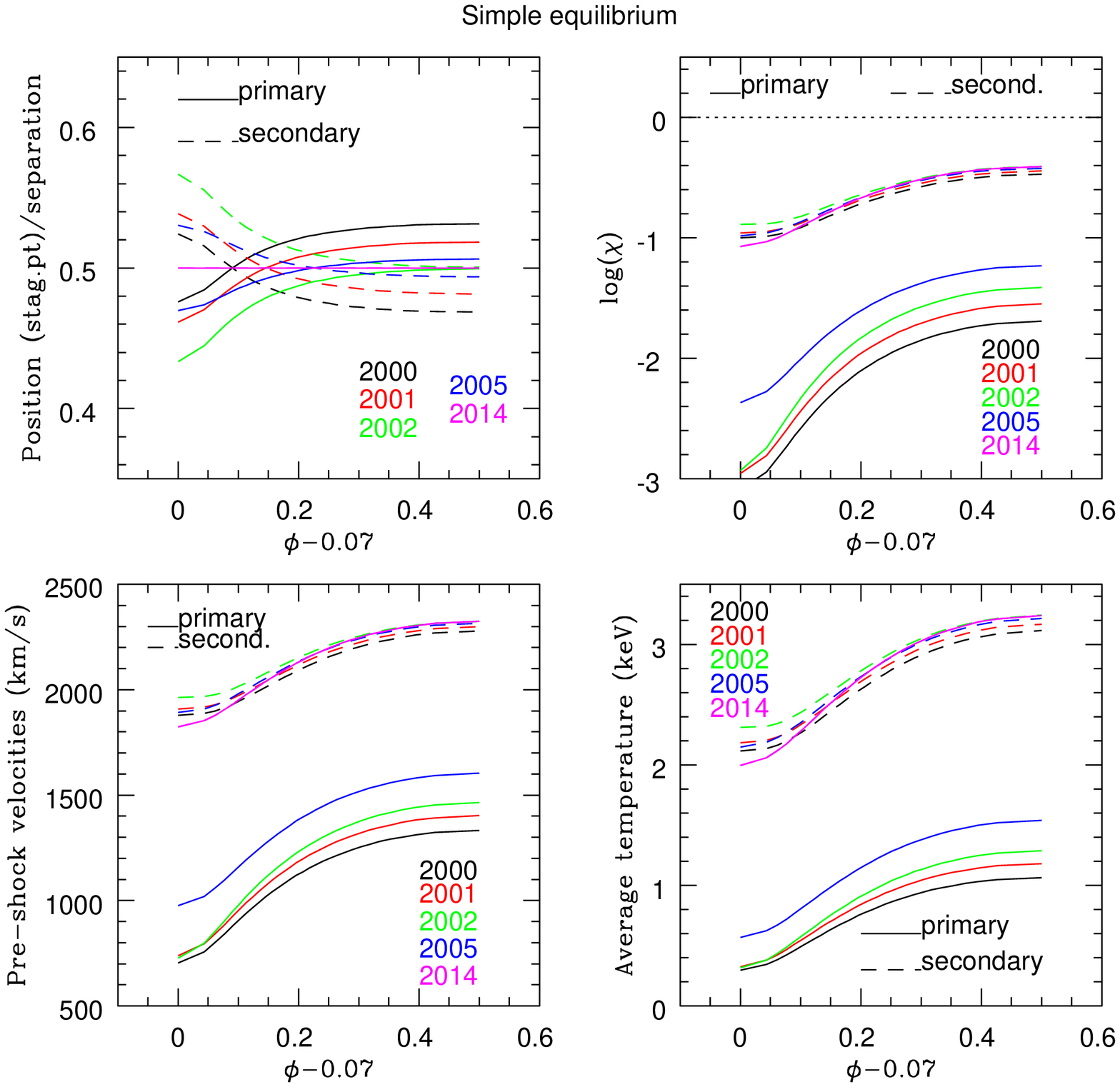}
\includegraphics[width=9cm]{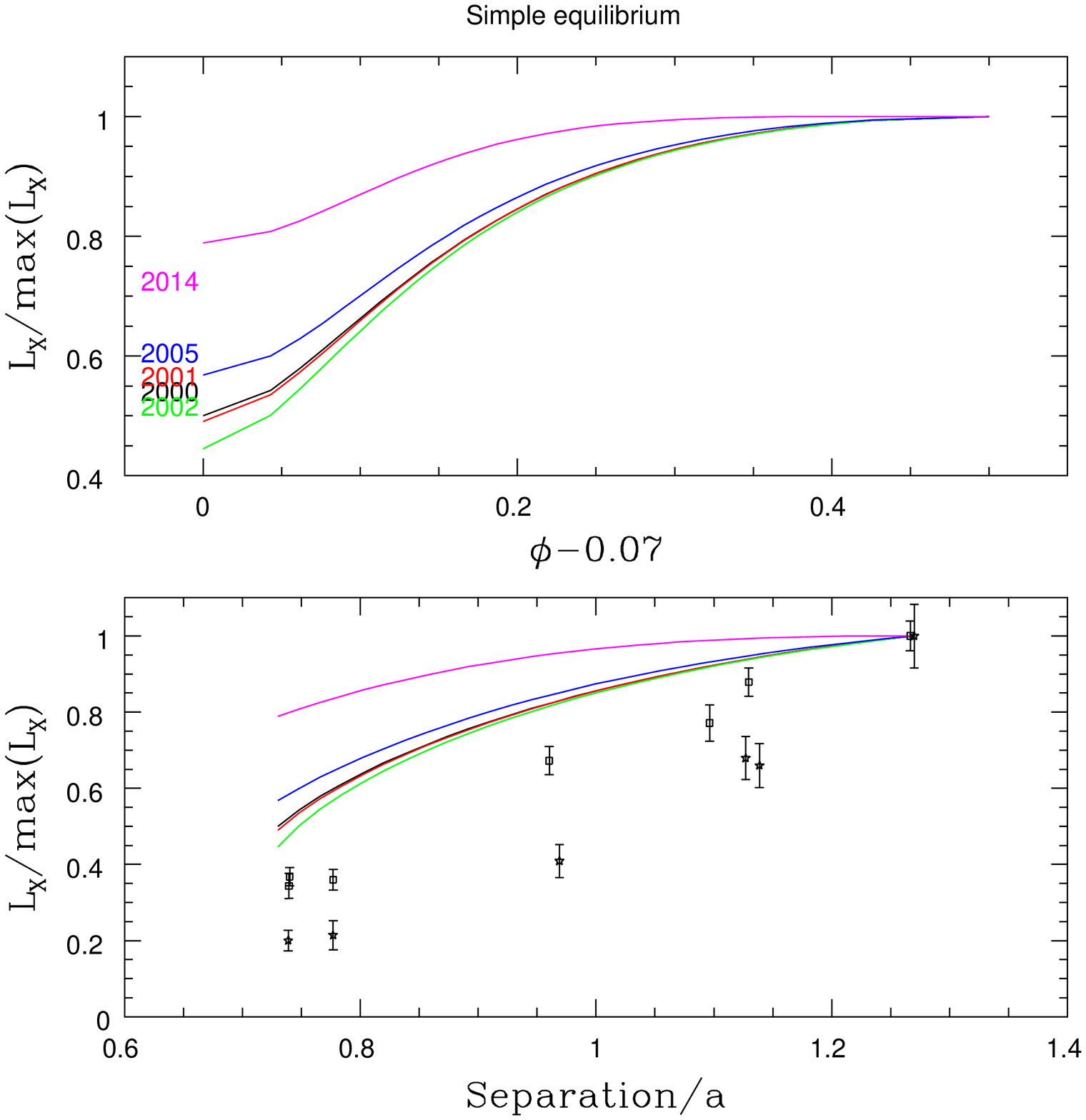}
\includegraphics[width=9cm]{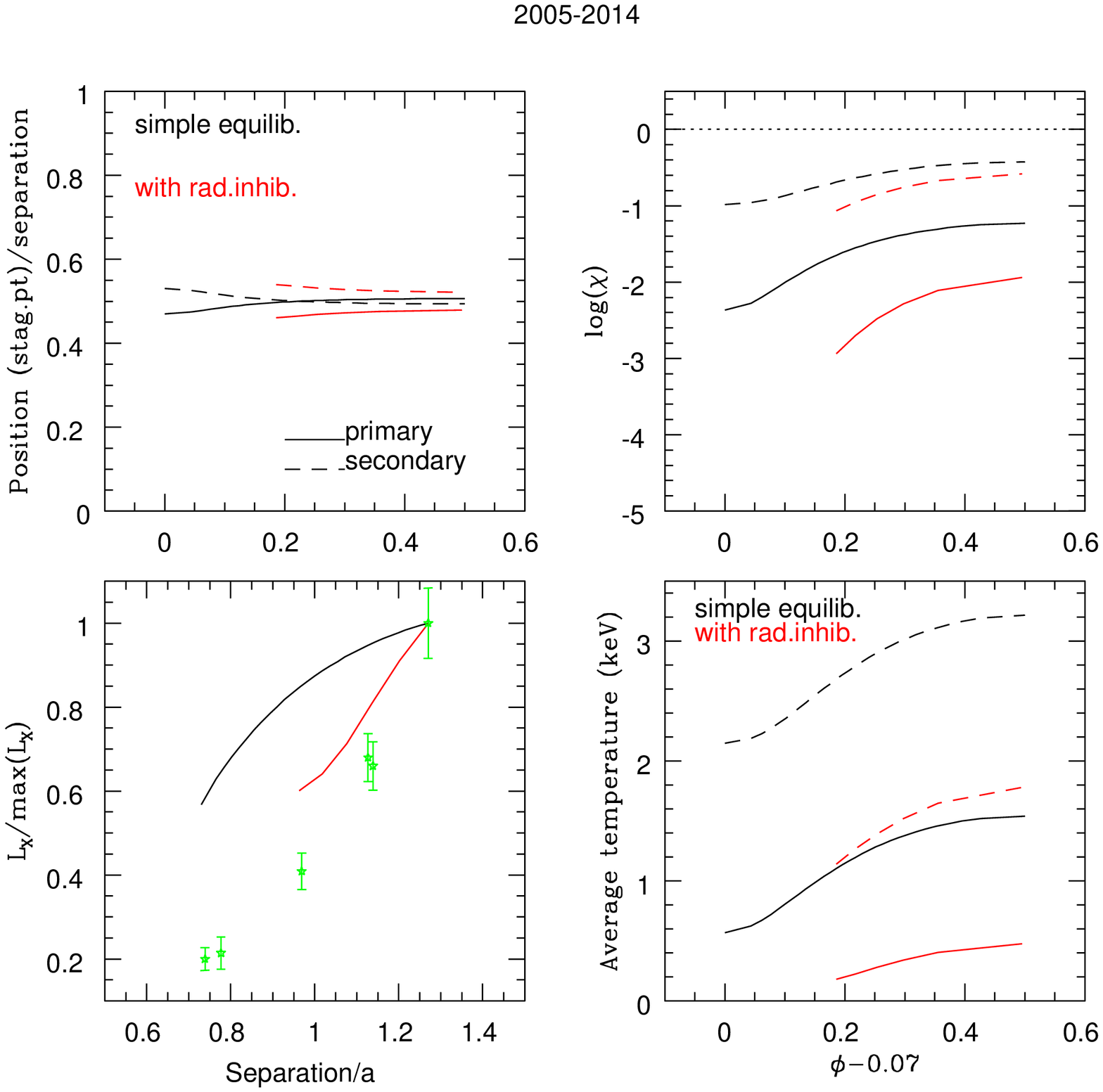}
\includegraphics[width=9cm]{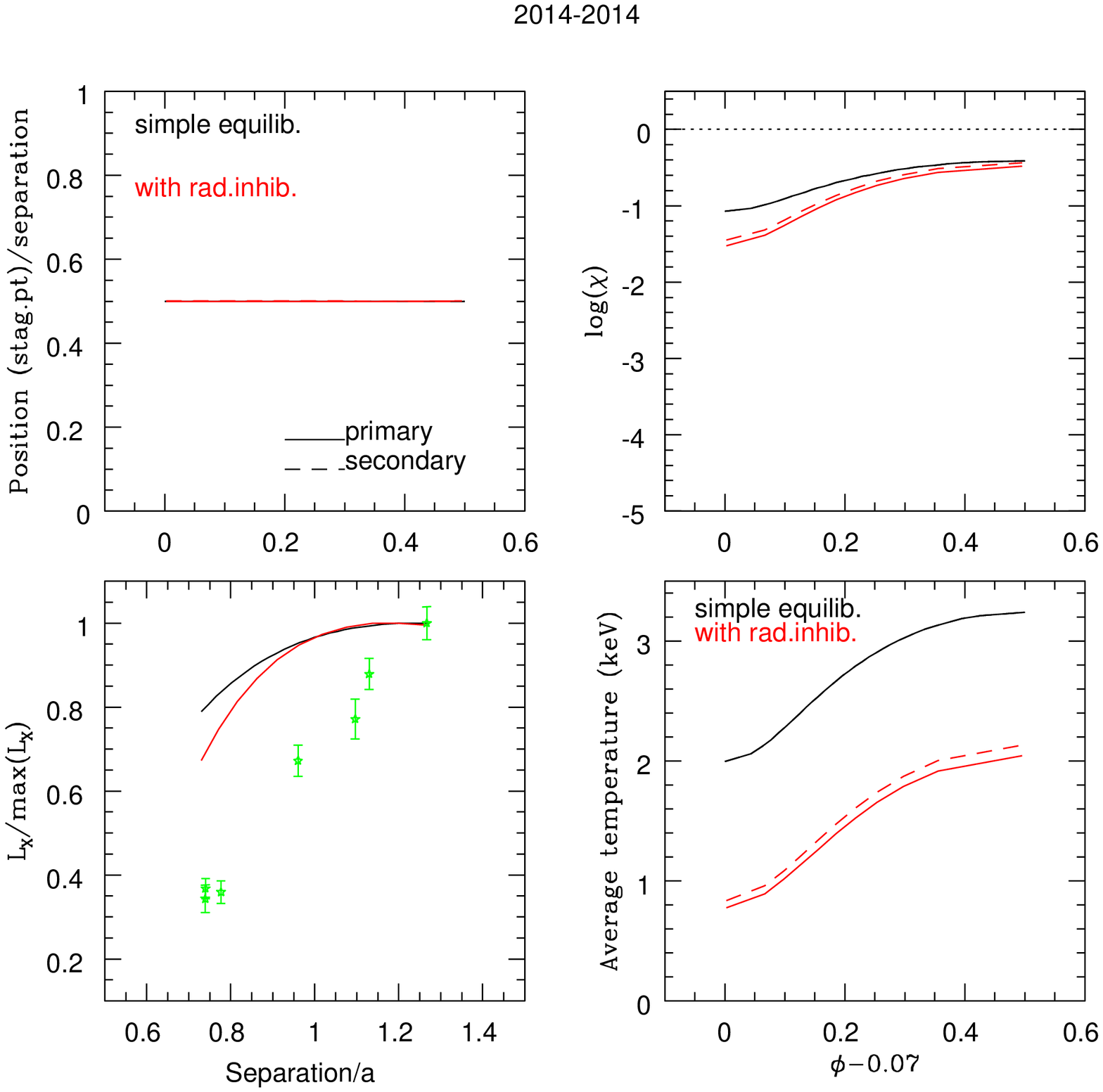}
\caption{{\it Top panels:} Results from simple equilibrium models considering the standard wind law and different properties of Star A (each year being identified by colors - see Table \ref{starA} for details). The values of the position of the stagnation point, the pre-shock velocities, the cooling parameters $\chi$,  the temperatures, and X-ray luminosities are shown as a function of phase. They are provided by solid or dashed lines if the primary (Star A) and the secondary (Star B), respectively, are separately considered. The total (i.e., summing contribution of both primary and secondary stars) X-ray luminosities shown on the right panels are normalized to the apastron values and compared to the MOS1+2 count rates, also normalized to apastron values (stars for 2000--2005 data, squares for 2016 data). {\it Bottom panels:} Comparison of results with or without inclusion of radiative inhibition, considering the properties of Star A in 2005 (left) and 2014 (right). The X-ray luminosities are again normalized to the apastron values, and compared to the associated MOS1+2 count rates, also normalized to apastron values (green symbols). In all panels, the x-axes report relative separation or phase, using $\phi-0.07$ to have 0 for periastron and 0.5 for apastron. }  
\label{models}
\end{figure*} 

\begin{figure}
\includegraphics[width=8cm]{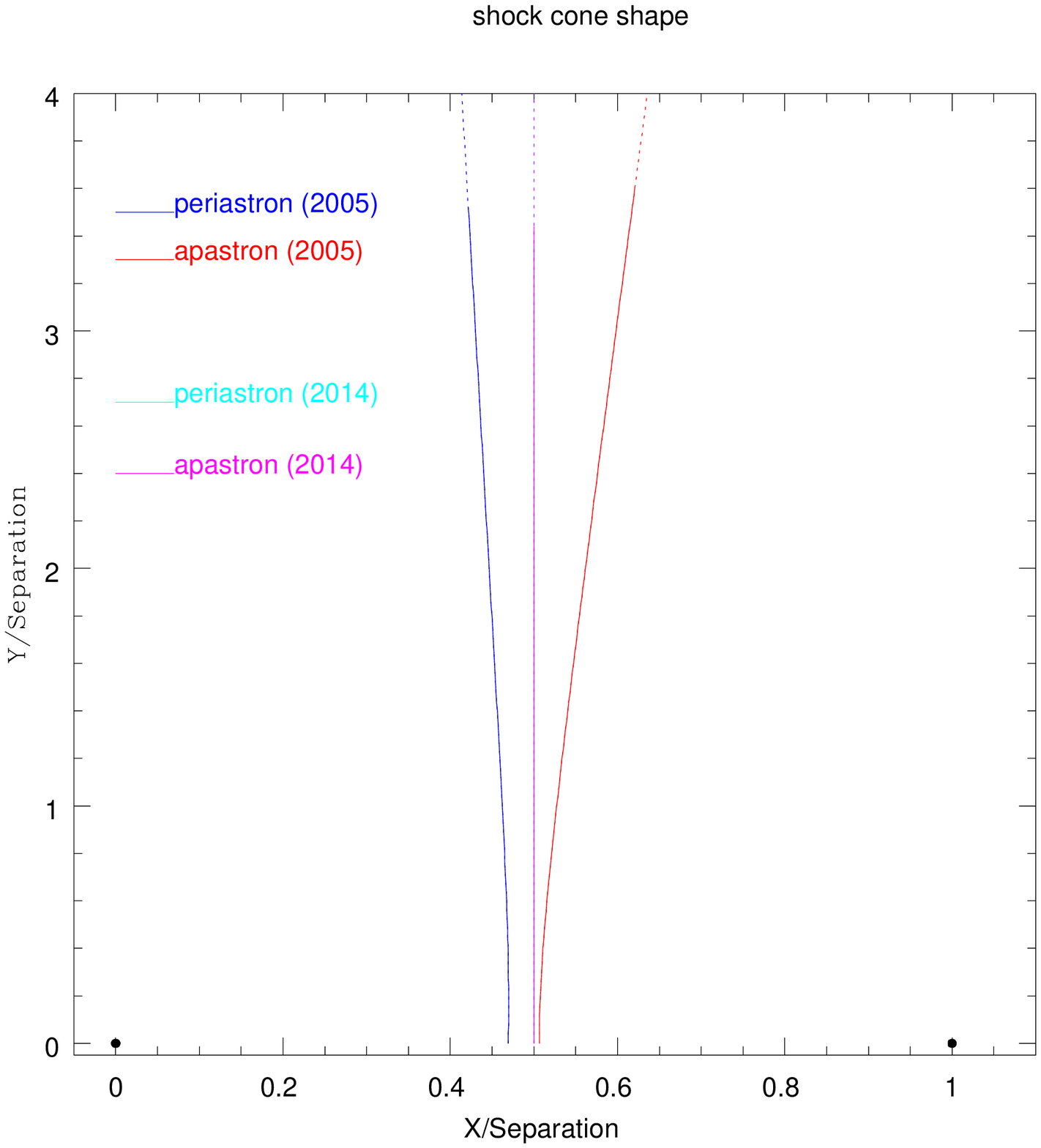}
\caption{Shape of the shock cone for 2005 and 2014, at apastron and periastron, with the part at $kT>0.2$\,keV shown as solid lines. The axes are distances divided by the separation at the phase considered; by convention, the two stars lie at (0,0) and (1,0). }  
\label{cshape}
\end{figure}

\vspace*{7mm}

To further interpret these phase-locked observations in the framework of a wind-wind collision in the A+B system, it is important to examine what is predicted by models. The X-ray flux produced by the wind-wind collision is some fraction of the wind kinetic luminosity $0.5 \dot M v^2$ \citep{ste92}, hence the precise knowledge of the wind properties is crucial. Indeed, the change in the wind properties of Star A from a slow eruptive phase to a fast, lower-density phase could produce an observable effect on the X-ray emission. The stellar parameters of Star A were derived by comparing optical/UV spectra with synthetic spectra computed with model atmosphere codes. To this aim, $HST$ data taken in 2000, 2002, 2009, and 2014 during the eclipse of Star B by Star A ($\phi\sim0$) were used. The fitting results using CMFGEN are reported by \citet{geo11} and Hillier et al. (in prep); they are summarized in Table \ref{starA}. Note that the larger wind terminal velocity in 2014 does not come from the CMFGEN fit itself but from the observed extension of the P Cygni profiles in the UV spectrum. The parameters fitted by \citet{she16}, using a subset of these observations and a different model atmosphere code, agree well with those reported by \citet{geo11} for the same data. Our X-ray observations were taken in 2000, 2001, 2005, 2013, and 2016: to get the properties of Star A in 2013 and 2016, we assume that it has not changed much since 2013, and we use the parameters from 2014; to get those for 2001 and 2005, we perform a simple interpolation (see italics in Table \ref{starA}). 

Contrary to Star A, Star B is assumed to have remained relatively stable over long timescales. In their Table 2, \citet{she16} list the properties derived for Star B, which agree well with those found for Star A in 2014 (Table \ref{starA}, Hillier et al., in prep), confirming the similarity of the two stars presented by \citet{koe14}. Furthermore, the CMFGEN analysis results of a $HST$ spectrum taken in 2016 when Star B eclipsed Star A ($\phi\sim0.36$) are also consistent with Star B having similar parameters as Star A in 2014 (Hillier et al., in prep). Therefore, we adopted the 2014 parameters of Star A for Star B. We finally note that Hillier et al. (in prep) do not find any evidence for a difference in the wind clumping factor $f$ between epochs for Star A or between Star A and Star B - a value of 0.1 is here adopted in all cases.

With these parameters, if we assume in addition that the winds collide at their terminal speeds, we find that the wind kinetic luminosity of Star A has not much changed since 2000, while the X-ray flux increased by a factor $\sim$2.5. At the same time, the wind momentum ratio $\dot M_Av_{\infty,A}/\dot M_Bv_{\infty,B}$ declined from 1.5 to 1.0, so the opening angle of the wind-wind collision cone should have increased (to near 180$^{\circ}$), while observations rather indicate similar lightcurve shapes over time. Finally, the wind density of Star A at a given position ($\rho\propto \dot M/v_{\infty}$) decreased by a factor of 3.3 between 2000 and 2014, while observations suggest a potential increase of absorption in recent years. From these numbers, it appears that the changing high-energy emission of \hd\ cannot be explained by these first, rough estimates.

In such close systems, it is however common that winds have not yet reached their terminal velocity before colliding, and the radiation from the companion may reduce the wind speed. Therefore, we performed next a more detailed calculation, based on the orbital parameters of \hd\ \citep{ste97,koe14}. For each orbital phase, we calculate the position of the stagnation point, where the wind momenta $\dot M v$ equilibrate along the line joining the two stars, considering two cases for the wind acceleration: the standard wind acceleration law given by $v=v_{\infty}[1-R_*/r]$ and a full radiative-driving calculation \citep{cas75,ste94,par13} taking the flux of both stars into account. Using the derived wind velocities, we then estimate the physical properties of the post-shock plasma of each wind, notably its temperature ($kT=3 mv^2/16$) and its intrinsic X-ray luminosity \citep[from the shocked wind formulation in Sect. 3.2 of][]{zab11} - this is an intrinsic value, i.e., without considering any wind absorption, hence it is representative of the hard emission. While a full hydrodynamic simulation (which is beyond the scope of this paper) is certainly needed, this simple modeling may clarify the general trends for the wind-wind collision properties, to be checked against data. 

Results of this modeling are shown in Fig. \ref{models}. The panels in the top two rows compare the results of the simple equilibrium (using the standard wind law) for different Star A parameters, while the panels of the bottom two rows compare the modeling results with and without radiative inhibition for the cases of 2005 (left) and 2014 (right). The equilibrium position always remains close to the mid-point (or half separation), as shown in the upper left panels of each part of Fig. \ref{models}. The collapse of the collision zone onto one of the star is thus avoided, unlike what is observed in some other eccentric binaries (e.g. $\eta$\,Car, \citealt{ham14} , WR21a \citealt{gos16}, HD\,166734 \citealt{naz17}). In addition, the collision always remains radiative in nature ($\chi<1$, see notably the middle top panel of Fig. \ref{models}), as derived from observations. 

We can also reproduce, at least qualitatively, the observed correlation between hard flux and separation (e.g. rightmost panel in second row of Fig. \ref{models}). A slightly better agreement is even obtained if radiative inhibition is included (last row of Fig. \ref{models}). In such radiative systems, the maximum occurs at apastron rather than at periastron, despite the decrease in wind density, because the impact of higher pre-shock wind speeds. Indeed, the wind-wind collision occurs as the winds still accelerate and they reach larger speeds when they are further apart, i.e. at apastron. Moreover, the wind braking effect due to the companion radiation is less strong when stars are more separated, reinforcing the increase in wind speed at apastron. However, such larger wind speeds at apastron also imply a larger temperature for the X-ray emitting plasma, while the observations rather suggest a decrease of temperature at that phase.

Finally, predictions for 2000--2005 are quite similar (top panels of Fig. \ref{models}), again as observed, but the modeling also predicts little change in the total X-ray brightness between 2005 and 2016. In addition, the predicted temperature (e.g. middle panel of the second row in Fig. \ref{models}) is twice as large for the primary wind in recent years because of its increased wind speed. This could explain the recent hardening, though such a large change in temperature is not obvious in spectra (fits rather favor an absorption increase). 

Since observations suggest the wind-wind collision zone to be large, we also calculated wind momenta equilibrium outside of the line-of-centers (without radiative inhibition). We derived the local temperature at these positions from the usual formula $kT=3mv^2/16$, where $v$ is the velocity component perpendicular to the shock \citep{ste92}. Considering X-rays to be emitted if $kT>0.2$\,keV, we found that the size of the emission regions are similar in 2005 and 2014 (Fig. \ref{cshape}), suggesting again similar flux levels at different epochs. 

\vspace*{7mm}

In summary, the change in Star A's mass-loss properties {\it alone} cannot explain the increased X-ray emission of \hd\ in 2016. However, the X-ray properties also depend on how the shocked stellar wind material cools as it flows away from the stars. Cooling can occur either by adiabatic expansion or via radiation. A characteristic measure of the relative importance of both processes is expressed by the cooling parameter $\chi =t_{cool}/t_{escape} \approx v^4 d/\dot M$ \citep[with the wind speed $v$ expressed in $10^8$\,cm\,s$^{-1}$, the separation $d$ in $10^{12}$\,cm and the mass-loss-rate in $10^{-7}$\,M$_{\odot}$\,yr$^{-1}$, see][]{ste92}. As stated above, the wind-wind collision in \hd\ appears to remain radiative (i.e. $\chi<1$) throughout the orbital cycle but the value of $\chi$ for Star A in 2000--2005 is much smaller than in 2016, by about an order of magnitude: in models using the standard wind law, $\chi$ at the stagnation point goes from 0.004 at periastron to 0.06 at apastron in 2005 and from 0.08 to 0.39 in 2016 (Fig. \ref{models}). These values depend of course on the chosen stellar parameters, notably wind clumping which affects the derived mass-loss rate. Using a lower wind clumping factor $f\sim 0.025$, as in \citet{geo11}, would lead to mass-loss rates smaller by a factor of two, hence to larger cooling parameters since $\chi\propto 1/\dot M$. A lower $f$ value would thus lead to an overall increase in $\chi$ but would not modify its trend. Thus radiative cooling a decade ago would have been even more dominant than in 2016. 

A recent theoretical study by \citet{kee14} has shown that the X-ray luminosity from highly radiative wind-wind collisions should be weaker than predictions from analytical models such as those used above. This is due to the onset of strong non-linear thin-shell instabilities in the radiatively cooling gas. In this case, the shocks become more oblique and since the post-shock temperature depends on the square of the velocity component perpendicular to the shock surface, the post-shock plasma is expected to be cooler. This theoretical study thus provides a possible explanation for the secular behavior of \hd\ since in 2000--2005, when the collision was more strongly radiatively cooled, the X-ray emission appeared both fainter and softer than in 2016. This tentative conclusion needs however to be confirmed by detailed 3D hydrodynamical simulations specifically dedicated to \hd.

\section{Conclusion} 
Two decades after the eruption in \hd, we have obtained new X-ray observations of the system. Surprisingly, the new data reveal a large increase of the X-ray brightness (factor of $\sim$2.5) at all phases, coupled to a hardening of the emission. This is an unprecedented feature for colliding wind binaries. However, the lightcurve {\it shape} has not changed significantly, with a tight linear correlation between the X-ray flux and the separation. The new data also further helped us refining our view of the phase-locked properties, enlightening the absence of eclipses and an increase of absorption at apastron coupled to a decrease in temperature at that phase. 

Simple analytical models of the collision, based on the known stellar and orbital parameters, explain the flux variation with separation, at least qualitatively, although they fail to reproduce some observed features - in particular, a larger temperature is rather expected at apastron. In these models, the change in stellar parameters of Star A appears consistent with the recent hardening (through an increase of plasma temperature), but the mean X-ray flux is predicted to remain similar whatever the epoch under consideration. The long-term brightness change could however be related to a recent theoretical suggestion of increased strength of thin-shell instabilities in highly radiative shocks. If confirmed by dedicated modeling, this would once more underline the exceptional astrophysical interest of the \hd\ system.

\acknowledgments
YN and GR acknowledge support from the Fonds National de la Recherche Scientifique (Belgium), the Communaut\'e Fran\c caise de Belgique, the PRODEX \xmm\ and Integral contracts, and an ARC grant for concerted research actions financed by the French community of Belgium (Wallonia-Brussels Federation). GK acknowledges support from CONACYT grant 252499. D.J.H. acknowledges support from Grants HST-GO-13373.001-A and HST-GO-14476.002-A. ADS and CDS were used for preparing this document. 


\end{document}